\documentclass[fleqn,usenatbib]{mnras}
\usepackage{newtxtext,newtxmath}
\usepackage[T1]{fontenc}
\usepackage{amsmath}
\usepackage[utf8]{inputenc}
\usepackage{graphicx}
\usepackage{float}
\usepackage{caption}
\usepackage{subcaption}
\usepackage[version=4]{mhchem} 
\usepackage{multirow}
\usepackage{verbatim}
\usepackage[normalem]{ulem}

\newcommand*{\ccfobs}{\ensuremath{\mathrm{CCF}_{\mathrm{obs}}}~}
\newcommand*{\ccfinj}{\ensuremath{\mathrm{CCF}_{\mathrm{inj}}}~}
\newcommand*{\deltaccf}{$\Delta$CCF~}
\newcommand*{\ccfobsa}{\ensuremath{\mathrm{CCF}_{\mathrm{obs}}}}
\newcommand*{\ccfinja}{\ensuremath{\mathrm{CCF}_{\mathrm{inj}}}}
\newcommand*{\deltaccfa}{$\Delta$CCF}
\newcommand*{\kp}{$K_{\mathrm{p}}$~}
\newcommand*{\vsys}{$V_{\mathrm{sys}}$~}
\newcommand*{\vwind}{$V_{\mathrm{wind}}$~}
\newcommand*{\vbary}{$V_{\mathrm{bary}}$~}
\newcommand*{\vp}{$V_{\mathrm{p}}$~}
\newcommand*{\kms}{$\mathrm{km~s^{-1}}$~}
\newcommand*{\kmsa}{$\mathrm{km~s^{-1}}$}

\title[Robustness in High-Resolution Spectroscopy]{Robustness Measures for Molecular Detections using High-Resolution Transmission Spectroscopy of Exoplanets}

\author[Cheverall et al.]{
Connor J. Cheverall,$^{1}$\thanks{E-mail: cjc218@ast.cam.ac.uk}
Nikku Madhusudhan,$^{1}$\thanks{E-mail: nmadhu@ast.cam.ac.uk}
Måns Holmberg$^{1}$
\\
$^{1}$Institute of Astronomy, University of Cambridge, Madingley Road, Cambridge CB3 0HA, UK\\
}

\date{Accepted 17 February 2023. Received 23 January 2023; in original form 18 August 2022.}
\pubyear{2023}
\begin{document}
\label{firstpage}
\pagerange{\pageref{firstpage}--\pageref{lastpage}}
\maketitle

\begin{abstract}
Ground-based high-resolution transmission spectroscopy has emerged as a promising technique for detecting chemicals in transiting exoplanetary atmospheres. Despite chemical inferences in several exoplanets and previous robustness studies, a robust and consistent detrending method to remove telluric and stellar features from transmission spectra has yet to be agreed upon. In this work we investigate the robustness of metrics used to optimise PCA-based detrending for high-resolution transmission spectra of exoplanets in the near-infrared. As a case study, we consider observations of the hot Jupiter HD~189733~b obtained using the CARMENES spectrograph on the 3.5~m CAHA telescope. We confirm that optimising the detrending parameters to maximise the S/N of a cross-correlation signal in the presence of noise has the potential to bias the detection significance at the planetary velocity of optimisation. However, we find that optimisation using the difference between a signal-injected cross-correlation function and the direct cross-correlation function (CCF) is more robust against over-optimisation of noise and spurious signals. We additionally examine the robustness of weighting the contribution of each order to the final CCF, and of S/N calculations. Using a prescribed robust methodology, we confirm \ce{H2O} in the atmosphere of HD~189733~b (S/N~=~6.1). We then investigate two further case studies, of exoplanets HD~209458~b and WASP-76~b, confirming \ce{OH} in the atmosphere of WASP-76 b (S/N~=~4.7), and demonstrating how non-robust methods may induce false positive or inflated detections. Our findings pave the way towards a robust framework for homogeneous characterisation of exoplanetary atmospheres using high-resolution transmission spectroscopy in the near-infrared.\\
\end{abstract}


\section{Introduction}

Thousands of exoplanets have been discovered to date. The first confirmed exoplanet orbiting a Sun-like star was observed in 1995 with the discovery of the hot Jupiter 51 Peg b \citep{mayor_jupiter-mass_1995}. Since then over 5000 confirmed exoplanets have been discovered\footnote{https://exoplanets.nasa.gov/}. We now know that the occurrence frequency of planets is high, approaching one per star \citep{fressin_false_2013, fulton_california-_2017}. Our accelerating ability to detect and subsequently characterise exoplanets is due to the rapidly improving technology, instrumentation and analysis capabilities available, and this will continue into the future. The field of exoplanets is therefore one of the most active and fast-paced frontiers in astrophysics.

Exoplanets are hugely diverse in terms of their orbital parameters, bulk parameters (masses, radii, equilibrium temperatures), internal structures, formation conditions and evolution histories. Their atmospheres span a wide range of chemical compositions and temperature profiles, with various chemical and physical processes at play \citep[e.g.][]{madhusudhan_exoplanetary_2014,madhusudhan_exoplanetary_2019,zhang2020,fortney2021}. Characterizing the atmospheres of exoplanets via the spectral signatures of the chemical species present allows us to constrain their diverse properties, contributing to an understanding of their physical processes and how they form. This in turn will enable us to learn more about the planets in our own solar system and their formation history.

Common molecules in exoplanetary atmospheres such as \ce{H2O}, \ce{CO}, \ce{CH4}, \ce{HCN}, \ce{CO2} and \ce{TiO}, and atoms such as \ce{Na} and \ce{K}, have strong absorption features in the optical and/or near-infrared (NIR) which may be seen in the transmission spectrum \citep{seager_theoretical_2000, sing_continuum_2016,madhusudhan_exoplanetary_2019}.  Analysis of the transmission spectrum can therefore constrain the composition of the atmosphere at the day-night terminator region. \cite{charbonneau_detection_2002} were the first to use transmission spectroscopy to characterise an exoplanetary atmosphere, using the Hubble Space Telescope (HST) to identify \ce{Na} in the atmosphere of the hot Jupiter HD 209458 b. It was not until 2008 that ground-based observations were first used to detect a chemical signature in an exoplanetary atmosphere, when \cite{redfield_sodium_2008} and \cite{snellen_ground-based_2008} made detections of \ce{Na} in the atmospheres of the hot Jupiters HD 189733 b and HD 209458 b, respectively.

In recent years, high-resolution transmission spectroscopy has emerged as one of the most successful techniques for detecting chemicals in transiting exoplanetary atmospheres \citep[e.g.][]{snellen_orbital_2010, wyttenbach_spectrally_2015, brogi_rotation_2016, brogi_exoplanet_2018, hoeijmakers_atomic_2018, alonso-floriano_multiple_2019, sanchez-lopez_water_2019, giacobbe_five_2021}. Whereas in low-resolution the molecular lines of different species may overlap, at high-resolution the signatures are more easily separated, giving more confident detections. High-resolution typically implies \textit{R} between $10^4$ and $10^5$ \citep{brogi_rotation_2016, van_sluijs_carbon_2022}, which is achieved by various NIR spectrographs on large (4-8m) ground-based telescopes. For a broadband absorber with lines of comparable strength, the signal-to-noise ratio (S/N) of the planetary signal increases with the square root of the number of lines observed, $\sqrt{N_{\mathrm{lines}}}$, so it is ideal to have a spectrograph of very high resolution and with a wide spectral coverage \citep{birkby_exoplanet_2018}.

This work focuses on the NIR wavelength range which contains strong spectral features of prominent molecules such as \ce{H2O}, \ce{CO}, \ce{CH4}, and \ce{HCN}, which are expected to be abundant in H$_2$-rich atmospheres \citep{moses_chemical_2013, madhusudhan_exoplanetary_2016}. There are a number of high-resolution spectrographs currently in use which cover the NIR, including CARMENES \citep{quirrenbach_carmenes_2016, quirrenbach_carmenes_2018}, CRIRES \citep{kaeufl_crires_2004, dorn_crires_2014}, GIANO \citep{oliva_giano-tng_2006, origlia_high_2014}, SPIRou \citep{the_spirou_team_spirou_2018, donati_spirou_2020}, IGRINS \citep{yuk_preliminary_2010, park_design_2014} and HDS/Subaru \citep{noguchi_high_2002}. The first detections achieved via high-resolution spectroscopy required spectographs mounted on 8 m class telescopes, such as CRIRES \citep{snellen_orbital_2010, brogi_signature_2012, birkby_detection_2013}. \cite{snellen_orbital_2010} were the first, using CRIRES to identify \ce{CO} in the transmission spectrum of HD 209458 b. More recently however, atmospheric characterisation of transiting exoplanets has been possible using spectographs mounted on 4 m class telescopes such as CARMENES \citep{alonso-floriano_multiple_2019, sanchez-lopez_water_2019}, GIANO \citep{brogi_exoplanet_2018, giacobbe_five_2021} and SPIRou \citep{boucher_characterizing_2021}. New instruments are continuously becoming available, increasing our capabilities even further and enabling the characterisation of more diverse exoplanetary atmospheres.

Close-in, and therefore strongly irradiated, gas giants called hot Jupiters are the most easily characterised, and most commonly observed, type of exoplanet. They are therefore the most common targets for high-resolution spectroscopy. Various chemical species such as \ce{CO} \citep{snellen_orbital_2010, brogi_signature_2012}, \ce{H2O} \citep{birkby_detection_2013, alonso-floriano_multiple_2019}, \ce{TiO} \citep{nugroho_high-resolution_2017}, \ce{HCN} \citep{hawker_evidence_2018, cabot_robustness_2019}, \ce{CH4} \citep{guilluy_exoplanet_2019}, \ce{NH3}, \ce{C2H2} \citep{giacobbe_five_2021, guilluy_gaps_2022, carleo_gaps_2022}, \ce{Fe}, \ce{Ti} and \ce{Ti+} \citep{hoeijmakers_atomic_2018} have been inferred in their atmospheres in both transmission and emission.

When observed using ground-based telescopes, NIR spectral lines produced by molecular species in the exoplanet's atmosphere are buried in stellar features and telluric contamination from the Earth's own atmosphere, both of which are orders of magnitude stronger \citep{sanchez-lopez_water_2019}. In order to access the planetary signal, the telluric and stellar lines first have to be removed in a process called detrending. This method typically makes use of the changing Doppler shift of the exoplanet's atmospheric spectrum as it transits in front of the host star. For a hot Jupiter, the orbital velocity of the planet is significantly greater than that of its host star. The planetary spectral lines are thus Doppler-shifted with a much greater amplitude than the stellar lines. Over a sufficient observation period, the planetary spectral lines will be subject to large Doppler shifts, whereas the telluric and stellar lines will remain comparatively stationary \citep{birkby_exoplanet_2018}. This allows us to separate the planetary spectral lines from those of the host star and the telluric absorption. Once the stellar and telluric lines have been removed, the planetary signal must be extracted from the noise. At high resolution, molecular features are resolved into a dense and unique collection of individual and separate lines, each with a very low S/N. Molecules can be detected by cross-correlating the observed high-resolution spectra, after detrending, with model atmospheric spectra of the planet \citep{snellen_orbital_2010, brogi_signature_2012, birkby_detection_2013}.

Detrending has proven to be the most challenging step in high-resolution spectroscopy. \cite{cabot_robustness_2019} previously investigated the robustness of a common detrending procedure in the context of NIR emission spectroscopy, and found that detrending parameters can potentially be overfit by optimising the detection significance at a single point in planetary velocity space. Doing so can lead to amplified or spurious detections at the expected planetary velocity. Whilst tests were proposed to aid in identifying such false positives, a robust detrending method to avoid them has yet to be established and agreed upon. Inhomogeneous methodologies across the literature can lead to inconsistencies in the quoted significance and robustness of detections \citep{spring_black_2022}, thereby hindering our ability to place tighter constraints on the compositional diversity of exoplanetary atmospheres. Consistent, robust and reproducible methods are therefore desirable.

In this work we investigate the robustness of molecular detections made using high-resolution transmission spectroscopy in the NIR. Our goal is not an exhaustive exploration of the model space aimed at detecting molecular species. Instead we are focused on assessing the relative robustness of molecular detections using different optimisations of a given detrending procedure for the same model template. In doing so, we aim to identify a robust recipe for detrending. Despite the greater transit depth in the NIR increasing the S/N of transmission spectra, telluric absorption is more severe at redder wavelengths meaning that detrending is more difficult. 

The paper is organised as follows. In Section \ref{methods} we introduce the general methodology by which a planetary signal can be extracted from the spectra, using observations of HD 189733 b as a case study. In Section \ref{optimisation_methods} we examine the robustness of different detrending optimisations from across the literature \citep{birkby_discovery_2017, alonso-floriano_multiple_2019, cabot_robustness_2019, sanchez-lopez_water_2019, giacobbe_five_2021, spring_black_2022, holmberg_first_2022}. Order weighting and other contributing factors in the determination of the detection S/N are discussed in Section \ref{factors}. Robust methods to achieve high confidence chemical detections in the atmospheres of exoplanets are used to analyse other datasets in Section \ref{case_studies}. Potentially spurious and inflated detections resulting from non-robust methods are also demonstrated. We summarise and discuss our results in Section \ref{summary}.

\section{Methods}\label{methods}

In this section we describe the main steps involved in analysing high-resolution spectroscopic observations of exoplanetary transmission spectra using the cross-correlation method. As a case study, we here focus on the hot Jupiter HD 189733 b and discuss the observations and the general approach to infer a chemical signature in its atmosphere.

\subsection{Observations}

In order to demonstrate our methods, as a test case we consider archival CARMENES observations of a transit of the hot Jupiter HD 189733 b on the night of 7th September 2017 \citep{alonso-floriano_multiple_2019}. HD 189733 b is an extensively studied hot Jupiter orbiting a bright K star (\textit{V} = 7.7 mag) \citep{bouchy_elodie_2005}. \ce{H2O} has previously been detected in its atmosphere using the same CARMENES observations as we use here \citep{alonso-floriano_multiple_2019}, as well as in other high-resolution transmission spectroscopy studies \citep{brogi_rotation_2016, brogi_exoplanet_2018}. \cite{brogi_rotation_2016} additionally found \ce{CO} in the atmosphere of this planet using CRIRES over a spectral range around 2.3 $\mu$m. CARMENES is mounted on the 3.5 m telescope at the Calar Alto Observatory and consists of two fiber-fed high-resolution spectrograph channels (VIS and NIR). In this work we only use the NIR channel, which observes a wavelength range of 960-1710 nm at a resolution of \textit{R} = 80400 over 28 spectral orders. Each channel is fed by two fibres: fibre A positioned on the target and fibre B on the sky to identify sky emission lines. The data consists of 46 observations (spanning planetary orbital phases -0.0348 $<\phi<$ 0.0359; $\phi=0$ corresponds to mid-transit), of which 24 are in transit. A median S/N of 134 was observed across the pixels. All observations were obtained with a constant exposure time of 198 s. The airmass increased from a minimum of 1.03 to a maximum of 1.32 over the course of the observing night. The system properties for HD 189733 b are given in Table \ref{hd189733b_parameters}.

\begin{table}
\begin{tabular}{ |p{1.0cm}||p{3.8cm}|p{2.5cm}|  }
 \hline
 Parameter & Value & Reference\\
 \hline
 P & $2.21857567\pm0.00000015$ d & \cite{agol_climate_2010}\\
 $\mathrm{T_{0}}$ & $2454279.436714\pm0.000015$ BJD & \cite{agol_climate_2010}\\
 $\mathrm{R_{star}}$ & $0.756\pm0.018$ $\mathrm{R_{\odot}}$ & \cite{torres_improved_2008}\\
 $\mathrm{R_{p}}$ & $1.138\pm0.027$ $\mathrm{R_{Jup}}$ & \cite{torres_improved_2008}\\
 \vsys & $-2.361\pm0.003$ \kms & \cite{bouchy_elodie_2005}\\
 a & $0.03120\pm0.00027$ au & \cite{triaud_rossiter-mclaughlin_2009}\\
 i & $85.71\pm0.024$ $^{\circ}$ & \cite{agol_climate_2010}\\
 $\mathrm{T_{14}}$ & $1.80\pm0.04$ hr & \cite{addison_minerva-australis_2019}\\
 \hline
\end{tabular} 
\caption{System properties of HD 189733 b, following  values used in \citet{alonso-floriano_multiple_2019}.}
\label{hd189733b_parameters}
\end{table}

\subsection{Data Cleaning, Normalisation and Calibration} \label{data_clean_norm_cal}

The pre-processed CARMENES data is publicly available, having been automatically reduced after observation using the dedicated pipeline CARACAL v2.10 \citep{zechmeister_flat-relative_2014, peck_carmenes_2016}. Throughout the analysis each spectral order is treated independently, until they are combined at the end. The first step is to remove bad pixels and outliers from the spectra. After cleaning each spectrum of poor quality pixels, each spectrum is first rescaled such that it's ninetieth percentile flux is unity. 5$\sigma$ outliers are then iteratively clipped from the time-series of each wavelength channel, with clipped pixels replaced using linear interpolation within that wavelength channel. The flux in each spectral order is then normalised by fitting a quadratic polynomial to the pseudo-continuum \citep{sanchez-lopez_water_2019}. Pixels identified by CARMENES fibre B as corresponding to sky emission lines are excluded from the normalisation fit. We remove orders 45-41 and 55-53 from this dataset due to their low S/N, and find that the CARMENES spectra require no further wavelength calibration.

\subsection{Detrending} \label{detrending}

The cleaned and normalised spectra are still dominated by telluric and stellar lines. In order to access the planetary signal, orders of magnitude weaker than such contaminants, they need to be removed (detrending). Principle component analysis (PCA) has been used to do this in several past studies \citep{de_kok_detection_2013, giacobbe_five_2021, holmberg_first_2022, van_sluijs_carbon_2022}. PCA finds and removes the common modes in the time-variation of each wavelength channel. As such, the quasi-static telluric and stellar features are removed by this process; during the course of the observations they vary only in depth, not significantly in wavelength, and so they produce common time-variations between wavelength channels. On the other hand, the planetary signal, with its changing Doppler shift, should mostly remain since it moves across different wavelength channels over the observing night. The planetary signal should therefore induce a minimal common time-variation between wavelength channels. An alternative algorithm known as SYSREM \citep{tamuz_correcting_2005, mazeh_sys-rem_2007} has also often been used instead. SYSREM allows for unequal uncertainties between pixels, and has been successfully used in a number of previous works \citep[e.g.][]{birkby_discovery_2017, nugroho_high-resolution_2017, hawker_evidence_2018, alonso-floriano_multiple_2019, sanchez-lopez_water_2019, cabot_robustness_2019, spring_black_2022}. We find minimal difference between residuals when detrending with each of PCA and SYSREM, and use PCA in this work.

\begin{figure}
    \centering
    \includegraphics[width=0.5\textwidth]{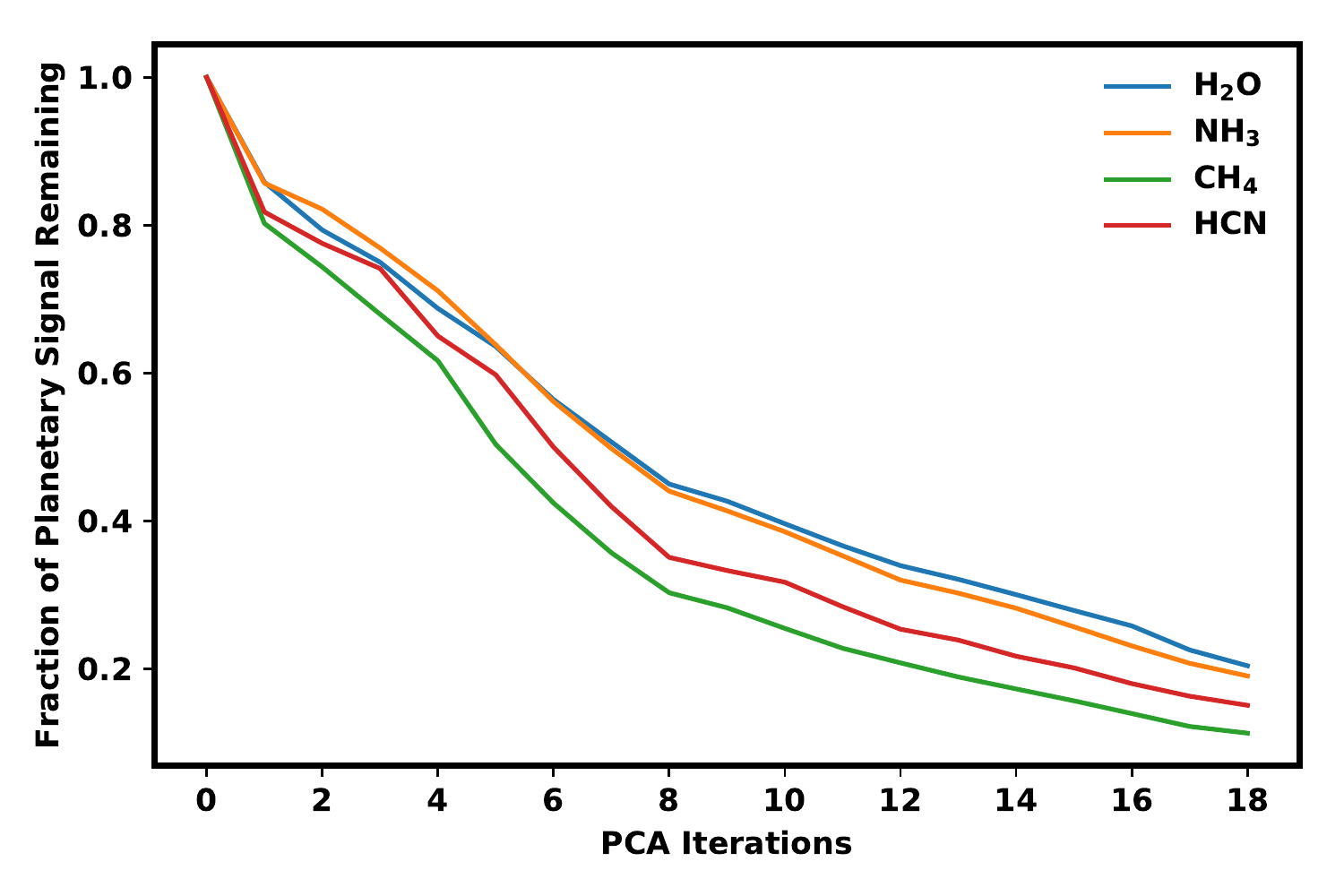} 
    \caption{The erosion of the total \deltaccf value of a planetary signal with each PCA iteration, for \ce{H2O}, \ce{NH3}, \ce{CH4} and \ce{HCN} models. In each case, $\lesssim$ 20$\%$ of the planetary signal, injected at the expected planetary velocity of HD 189733 b, remains after 18 iterations of PCA. Calculation of the total \deltaccf value is explained in Sections \ref{signal_extraction} and \ref{delta_method}.}
    \label{fig:signal_erosion}
\end{figure}

The number of PCA iterations applied refers to the number of principle components removed from the spectra. PCA iterations are applied to the spectra until, in principle, we are left with the continuum-normalized planetary spectrum embedded only in white noise \citep{birkby_exoplanet_2018}. Sufficient iterations of the detrending algorithm must be applied to remove the contaminants completely. As noted above, the planet signal should remain mostly intact assuming its change in radial velocity is sufficiently large. However, in reality detrending also erodes the planetary signal itself \citep{birkby_discovery_2017, sanchez-lopez_water_2019}. We test this by injecting a model planetary signal into the spectra and recovering it, as described in Section \ref{signal_extraction}, after different numbers of PCA iterations have been applied to detrend the spectra. We find that the planetary signal, injected at the expected planetary velocity, is degraded to some extent even after just one PCA iteration. In the case of this particular dataset, $\sim$15$\%$ of the planetary signal is lost after one PCA iteration. After 18 iterations less than 20$\%$ of the planetary signal remains for this dataset (Figure \ref{fig:signal_erosion}). We note however that this level of degradation could depend on the orbital properties of the planet. Therefore, there exists an optimum number of iterations which when applied to the spectra will retrieve the planetary signal with the greatest S/N. This optimum can vary between different spectral orders. Typically, it can be found by temporarily injecting a Doppler-shifted model planetary signal and maximising the retrieved detection significance at the desired location in velocity space \citep{birkby_discovery_2017, nugroho_high-resolution_2017, sanchez-lopez_water_2019}. The robustness of different such injection-based detrending optimisations is explored in more detail in Section \ref{optimisation_methods}.

\subsection{High-Resolution Model Spectra}

In order to cross-correlate with the data, we compute model templates for the transmission spectra of the hot Jupiters considered in this study. We model the transmission spectra using a variant of the AURA atmospheric modelling and retrieval code \citep{pinhas_retrieval_2018}. The model computes line-by-line radiative transfer in transmission geometry assuming a plane-parallel atmosphere in hydrostatic equilibrium. The atmospheric structure is computed over a pressure range of $10^{-7}$ - $100$ bar. The chemical composition and temperature structure are free parameters in the model. We generate the spectra considering one molecule at a time, assuming no clouds/hazes and an isothermal temperature profile. The spectra are computed at high resolution (\textit{R} $\gtrsim$ 10$^5$) over the CARMENES NIR spectral range, with opacity contributions due to prominent molecules expected in H$_2$-rich atmospheres over this temperature range (H$_2$O, CH$_4$, NH$_3$, HCN and OH) and assuming a nominal mixing ratio of 10$^{-4}$ for each molecule. We consider nominal isothermal temperature profiles at 1000 K, 1000 K and 2000 K for HD 189733 b, HD 209458 b, and WASP-76 b, respectively. The molecular cross-sections were obtained following the methods of \cite{gandhi_genesis_2017} using absorption line lists from the following sources: H$_2$O \citep{barber2006,rothman_hitemp_2010}, CH$_4$ \citep{yurchenko_exomol_2014}, NH$_3$ \citep{yurchenko_variationally_2011}, HCN  \citep{harris2006,  barber_exomol_2014}, and OH \citep{bernath2009,gordon_hitran2020_2022}. We also include collision-induced absorption from H$_2$-H$_2$ and H$_2$-He \citep{borysow1988,orton2007,abel2011,richard_new_2012}.

The model templates are separated into orders before undergoing the same normalisation as the observed spectra, as described in section \ref{data_clean_norm_cal}. We note that it is the relative depths and positions of the template absorption lines which matter in cross-correlation, and not their absolute depths \citep{sanchez-lopez_water_2019}. The model templates are then convolved with the point spread function of the instrument before cross-correlating with the detrended data.

\subsection{Signal Extraction} \label{signal_extraction}

Cross-correlating the detrended residuals with a model template allows us to combine the information from each line, such that we are able to extract a significant detection from the residuals \citep{snellen_orbital_2010, birkby_detection_2013, brogi_rotation_2016, birkby_exoplanet_2018}. A strong cross-correlation with a certain model indicates the presence of that molecule in the exoplanet's atmosphere.

For each spectral order independently, we start with the residual spectra at different phases. For each spectrum we first calculate the cross-correlation function (CCF) over a pre-determined velocity grid (-400 \kms to 400 \kms in steps of 1 \kmsa). To do so, we Doppler shift the model spectrum by each velocity in this grid and then cross-correlate it with the observed spectrum. This gives a cross-correlation value for each point in the velocity grid. Linear interpolation is used to project the Doppler-shifted model template onto the data wavelength grid \citep{sanchez-lopez_water_2019}. By repeating this over all phases we have for each order a CCF matrix in velocity and phase. The peak traces out the radial velocity of the planet with time. The mean is subtracted from each row, i.e. along the velocity axis at each phase, to remove broad variations between each spectrum \citep{alonso-floriano_multiple_2019, sanchez-lopez_water_2019}. All the order-wise CCF matrices are then summed to give a single CCF matrix for the entire spectral range. We subsequently shift the co-added CCF matrix into the planet-frame, for each point in a grid over planetary velocity space. 
Assuming a circular orbit, the planetary radial velocity \vp is given by
\begin{equation}\label{v_p}
V_{\mathrm{p}} = K_{\mathrm{p}} \sin (2\pi\phi) + V_{\mathrm{sys}} - V_{\mathrm{bary}} + V_{\mathrm{wind}}
\end{equation}
where \kp is the semi-amplitude of the planet's orbital motion, \vsys is the systemic velocity of the planetary system, \vbary is the barycentric velocity correction and \vwind accounts for any planetary atmospheric winds. The values of \vsys and \vbary are accurately known from the literature for each of the planets considered in this work. 

We explore a grid in \kp - \vwind space and obtain the total CCF at each point as follows. The planetary radial velocity at each point in this space, as calculated from equation (\ref{v_p}), is a function of phase. The rows of the CCF matrix, one for each phase, are shifted by the corresponding radial velocity. The total CCF is calculated by summing the cross-correlation values over time to give a one-dimensional distribution against planetary velocity. Whilst we use all spectra in detrending, only in-transit spectra are included in this addition. The separation of spectra into the in- and out-of-transit regimes can be done using the known orbital parameters of the system. To then obtain a detection significance at each point in \kp - \vwind space, the signal is taken as the value of the total CCF at zero velocity, whilst the noise is estimated by the standard deviation of the total CCF distribution away from this point. In estimating the noise we exclude velocities within $\pm$15 \kms to ensure that the measured signal does not influence the noise estimate.

For the correct point in \kp - \vwind space, the cross-correlation matrix will be shifted into the planet's true rest frame.  In this case, a peak in the CCF is obtained at zero velocity, maximising the detection significance. We therefore expect a high S/N peak at the planet's location in \kp - \vwind space if the atmosphere contains the chemical species present in the model template and our analysis is sufficiently robust. For HD 189733 b, the expected value for \kp is $152.5^{+1.3}_{-1.8}$ \kms \citep{brogi_rotation_2016}. Any offset from the expected systematic velocity may be attributed to atmospheric winds at the planetary terminator contributing an additional Doppler shift. Atmospheric winds have been constrained for a number of exoplanets in this way \citep{snellen_orbital_2010, brogi_rotation_2016, brogi_exoplanet_2018, alonso-floriano_multiple_2019, sanchez-lopez_water_2019}. The continued retrieval of a signal in only the out-of-transit spectra suggests that the signal may be spurious.

The Welch t-test \citep{welch_generalization_1947} is an alternative metric to quantify the detection significance \citep{birkby_discovery_2017, nugroho_high-resolution_2017, hawker_evidence_2018, alonso-floriano_multiple_2019, cabot_robustness_2019, sanchez-lopez_water_2019}. The shifted CCF matrix is split into two distributions: the `in-trail' distribution, covering the planet signal, and the `out-of-trail' distribution, which contains the cross-correlation noise. The Welch t-test is used to compare the two distributions and quantify the significance of the in-trail distribution's increased mean. In contrast to the S/N metric, this test's consideration of the standard deviation of each distribution means it may be less vulnerable to noisy pixels in the CCF falsely boosting the detection significance. However, \cite{cabot_robustness_2019} suggest that the Welch t-test may overestimate the confidence of detections due to oversampling, since correlations within the two distributions are typically not accounted for; also see \cite{collier_cameron_line-profile_2010}. Despite this issue potentially being solved if the aforementioned correlations are accounted for, the simpler S/N metric does not have this same problem, and is therefore more commonly used. We henceforth use the S/N metric in this work. We do however note that other factors can influence this metric. The effect of the explored velocity range on the noise estimate is investigated in Section \ref{vel_range}.

\section{Robustness of Detrending Optimisation Methods} \label{optimisation_methods}

Across the literature there are different methods for optimising the number of PCA iterations to apply in detrending. In this section we explore the robustness of some commonly used methods which involve the optimisation of detrending using an injected signal. We initially consider global detrending, where an equal number of PCA iterations is applied to each and every spectral order for a single night of observations. In the latter part of this section, we additionally consider optimisation of order-wise detrending.

\subsection{Direct CCF Optimisation} \label{injection_method}

In this method we optimise the detrending parameters based on the recovery of a synthetic signal injected into the data. The number of PCA iterations is selected to maximise the recovery of the signal which has been injected into the normalised in-transit spectra \citep{birkby_discovery_2017, nugroho_high-resolution_2017, alonso-floriano_multiple_2019, sanchez-lopez_water_2019}. For each PCA iteration, the detrended residuals are cross-correlated with the Doppler-shifted model template to derive the signal-injected CCF, which we refer to as \ccfinja. The iteration which returns the maximum S/N from \ccfinj at the injected planetary velocity is then selected for the detrending of the observed spectra. This optimisation method can also be done directly without first injecting a model signal \citep{alonso-floriano_multiple_2019, landman_detection_2021}. In this case, the iteration which optimises the S/N from the direct, or `observed', CCF, referred to as \ccfobsa, at the known planetary velocity is selected.

However, it has previously been suggested that such detrending optimisation methods are vulnerable to the overfitting of noise at the very point in planetary velocity space where we expect the signal, thereby falsely amplifying the detection significance \citep{cabot_robustness_2019}.

\subsection{Differential CCF Optimisation} \label{delta_method}

When optimising the number of PCA iterations, we aim to maximise the recovery of the planetary signal itself, rather than select an optimum iteration based on its amplification of noise into a more significant but biased detection. A less commonly used approach involves selecting the number of PCA iterations by optimising the S/N from a noise-subtracted CCF \citep{spring_black_2022, holmberg_first_2022}. Both \ccfobs and \ccfinj are found individually for each PCA iteration, as discussed in Section \ref{injection_method}. A differential CCF, \deltaccf \citep{brogi_rotation_2016, hoeijmakers_searching_2018}, can then be calculated for each iteration as:
\begin{equation}
\mathrm{\Delta CCF  =  CCF_{inj} - CCF_{obs}}.   
\end{equation}
When calculating the S/N from \deltaccfa, the signal is obtained from the \deltaccf matrix whereas the noise is estimated using \ccfobsa. Detrending parameters can then be selected to optimise this S/N at the expected planetary velocity. This approach allows us to optimise the detrending parameters on a model planetary signal with minimal noise at its location in planetary velocity space. Although information about residual noise around the injected planetary velocity is lost, we avoid the amplification of any noise which can falsely increase the detection significance. The extent to which this method may be more robust is investigated throughout this section.

\subsection{Comparison of Methods} \label{comparison}

To compare the robustness of the above optimisation methods, we use the previously reported detection of \ce{H2O} in observations of HD 189733 b \citep{alonso-floriano_multiple_2019} as a test case. Using the methods presented in Section \ref{methods}, we recover this detection of \ce{H2O} for a wide range of PCA iterations.

At this point, we do not consider the optimisation of the S/N from \ccfinja. Detrending parameters found by optimising the S/N from \ccfinj are dependent on the strength and structure of the injected model. There is no agreed upon injection strength in the literature, and different works use independently generated models, so results given by this method are inconsistent and difficult to reproduce. Conversely, we find that detrending parameters derived by optimising the S/N from \deltaccf are relatively independent of injected model strength. We later show that optimised detrending parameters found in this way show little variation across atmospheric models with the different chemical species that we consider. Since \ccfinj is similar to \ccfobs in the case of a weak injection and comparable to \deltaccf for a strong injection, we consider \ccfobs and \deltaccf as two extremes, from which conclusions about the performance and robustness of optimising the S/N from \ccfinj can be drawn. We therefore only compare the optimisations of the S/N from \ccfobs and \deltaccf in the remainder of this section.

\begin{figure}
    \centering
    \includegraphics[width=0.5\textwidth]{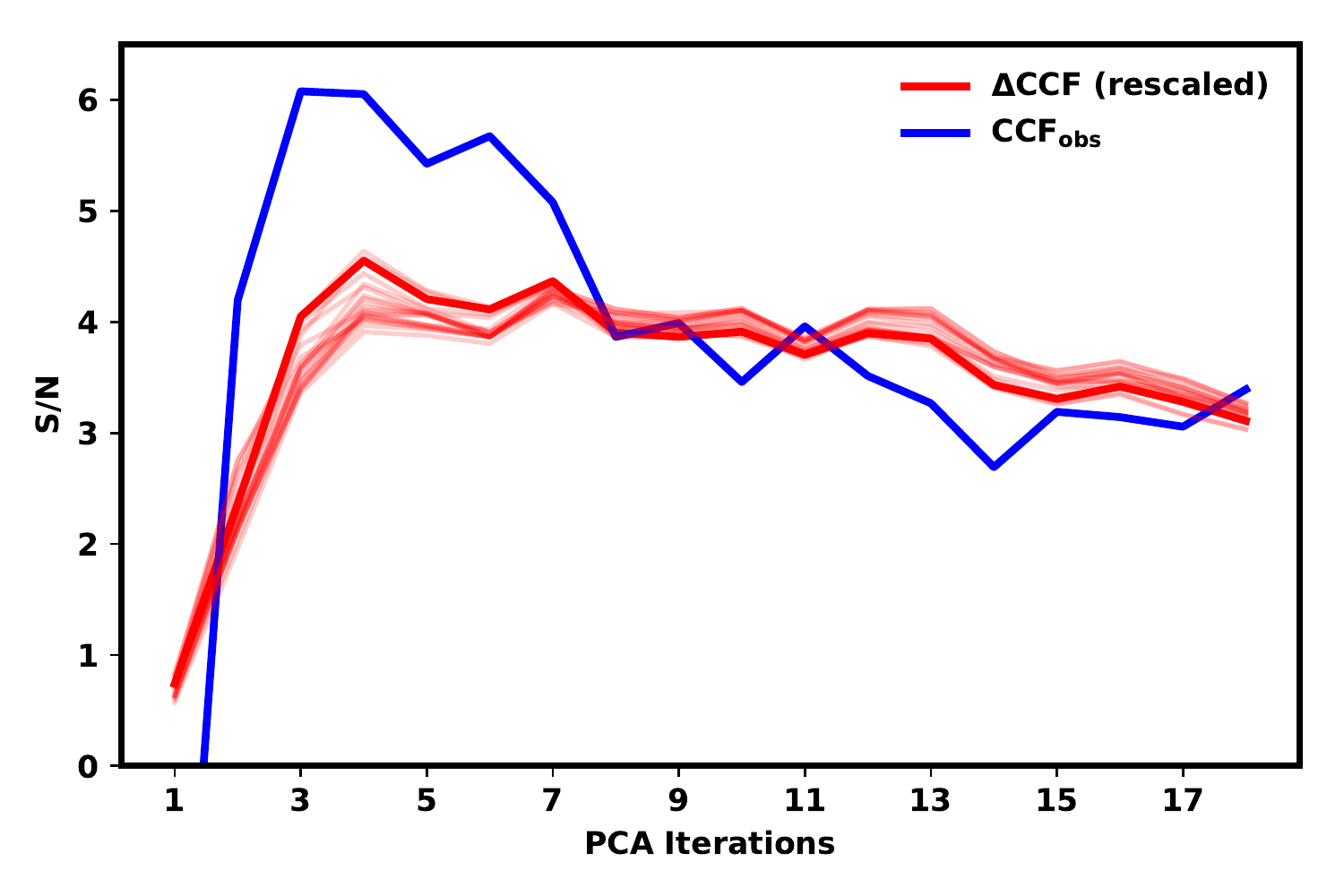} 
    \caption{The variation of retrieved S/N, from each of \ccfobs (blue) and \deltaccf (red), at the expected planetary velocity against the number of applied PCA iterations. An \ce{H2O} model is used, and iterations from 1 to 18 are considered. Faint red lines show the S/N from \deltaccf for different values of \vwind in the interval $\pm$50 \kmsa, for constant $K_{\mathrm{p}}$.}
    \label{fig:hd189733b_sn_pca}
\end{figure}

We begin by examining how the S/N at the expected planetary velocity, from each of \ccfobs and \deltaccfa, varies with the number of applied PCA iterations between 1 and 18 (Figure \ref{fig:hd189733b_sn_pca}). Whilst the shape of the \deltaccf S/N variation is reasonably invariant to the injection strength of the model, the absolute S/N values are approximately proportional to this strength. Since the absolute values are therefore arbitrary, we rescale the S/N from \deltaccf to have the same median as that from \ccfobs for $\ge$2 PCA iterations, such that the injected signal mimics the real signal. This rescaled S/N from \deltaccf demonstrates a fairly smooth variation with the number of applied PCA iterations, and may provide an estimate of the significance of the planetary signal. We find that the optimum PCA iteration is relatively independent of the strength of our injected model, an advantage over optimising the S/N from \ccfinj alone.

We observe that the variation in S/N from \ccfobs with the number of PCA iterations is noisier than that from \deltaccfa. If we assume that the rescaled S/N from \deltaccf is representative of the true planetary signal then noise in \ccfobs will give a greater than expected observed S/N for some iterations, and a lower than expected observed S/N for others.
If we optimise the S/N from \ccfobs in the detrending, then an iteration where noise components increase the detection significance will likely be selected. In other words, we could systematically inflate the detection significance by methodically selecting detrending parameters which amplify noise at the expected planetary velocity. This is further investigated throughout the remainder of this work.

\subsection{Optimisation Bias of Detrending Methods}

The extent to which the detection S/N is systematically increased due to amplified noise is here referred to as the bias. A detrending method is robust if the detection significance is unbiased, such that the expected S/N is not systemically increased or decreased. For example, a simple detrending method is to consistently across datasets apply an arbitrary and fixed number of PCA iterations. Although this sometimes will give a greater than expected S/N, it will also sometimes return a lower than expected S/N, creating a distribution of S/N values about the expected significance level. Such a detrending method may therefore be expected to be unbiased. A robust detrending optimisation method maximises the expected detection significance without inducing a bias.

\begin{figure}
     \centering
     \begin{subfigure}[h]{0.5\textwidth}
         \centering
         \includegraphics[width=\textwidth]{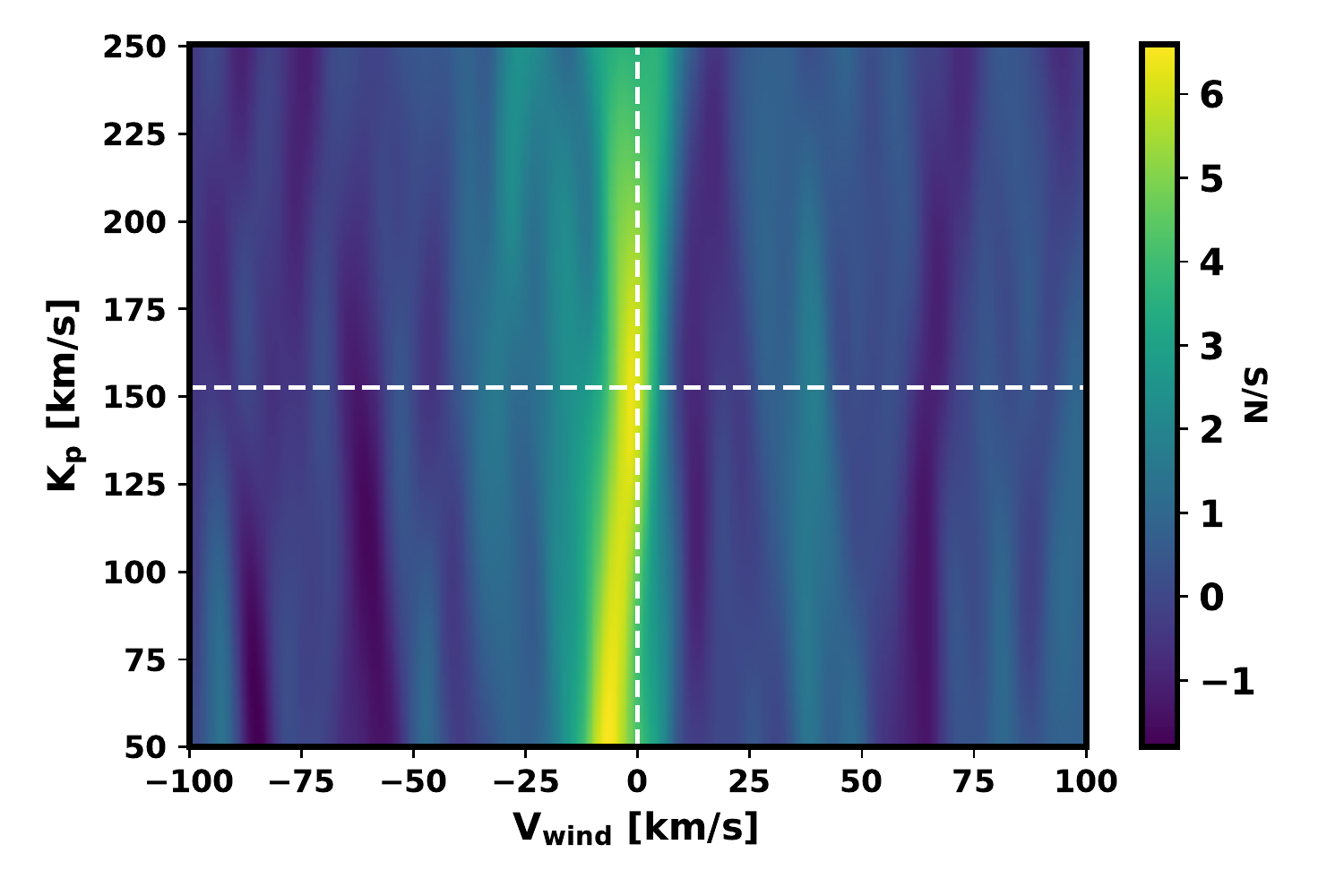}
         \caption{}
        \label{fig:hd189733b_h2o_4}
     \end{subfigure}
     \hfill
     \begin{subfigure}[h]{0.5\textwidth}
         \centering
         \includegraphics[width=\textwidth]{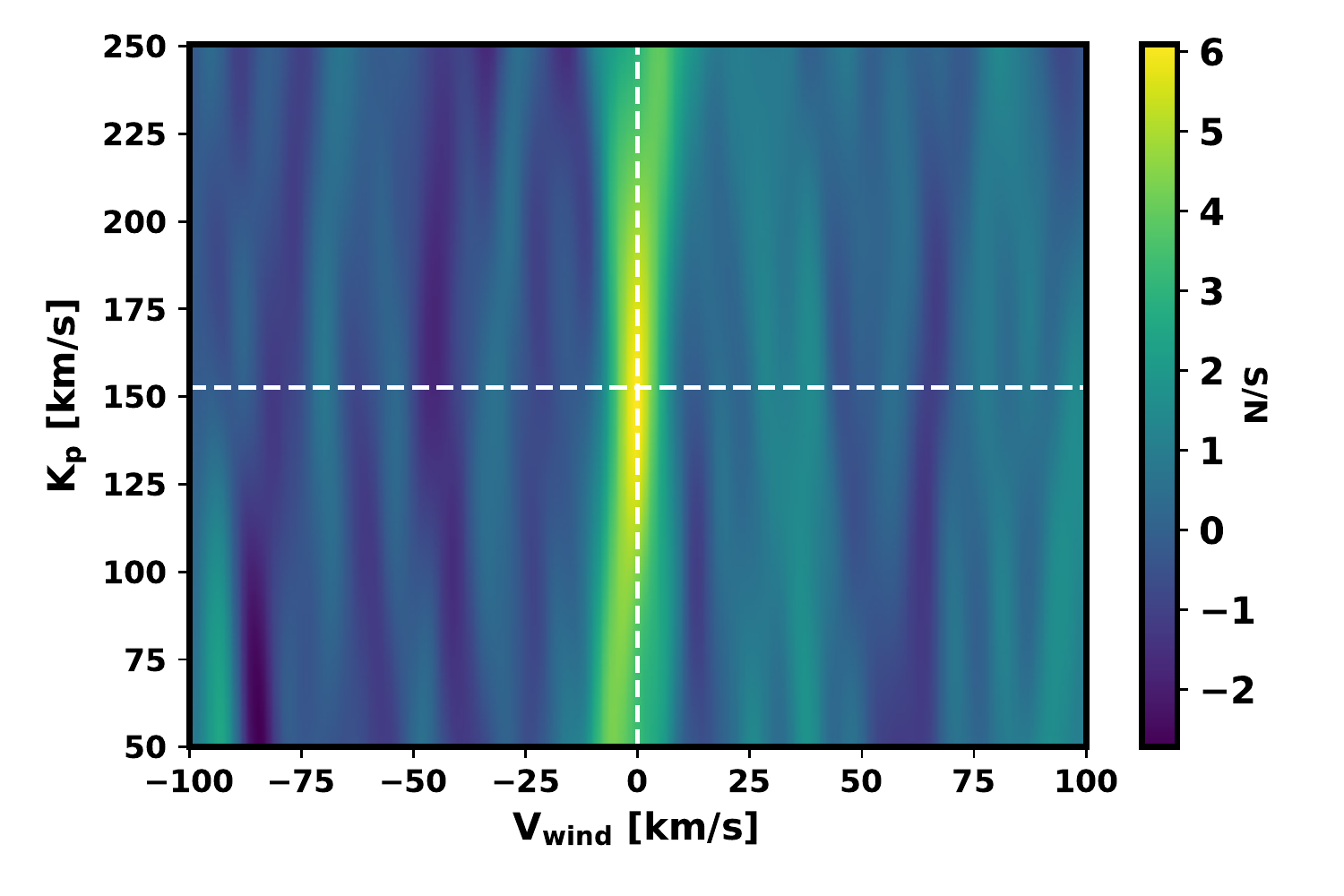}
         \caption{}
        \label{fig:hd189733b_h2o_7}
     \end{subfigure}
 \caption{S/N maps showing the retrieval of \ce{H2O} signals in the atmosphere of HD 189733 b via the detrending optimisation of the S/N from \ccfobs and \deltaccfa, respectively. Panel (a): applying 3 PCA iterations to optimise the S/N from \ccfobs retrieves a detection for \ce{H2O} with a S/N of 6.4. The median S/N across velocity space away from the expected planetary velocity (|\vwind| > 10 \kmsa) is 0.2. Panel (b): applying 4 PCA iterations to optimise the S/N from \deltaccf retrieves a detection for \ce{H2O} with a S/N of 6.1. The median S/N away from the expected planetary velocity is 0.0.}
 \label{fig:hd189733b_h2o}
\end{figure}

In Figure \ref{fig:hd189733b_sn_pca}, the S/N from \ccfobs is optimised by 3 PCA iterations, whereas the S/N from \deltaccf is optimised by 4 iterations. We detrend our spectra globally in each of these cases; the results obtained are shown in Figure \ref{fig:hd189733b_h2o}, with retrieved S/N values at the expected planetary velocity of 6.4 and 6.1, respectively. Here we show a case where the optimisation of detrending using the \deltaccf metric provides a number of PCA iterations very close to that when using \ccfobsa; greater by just one iteration. This leads to consistent values for the detection S/N. Generally however, the number of PCA iterations, and the subsequent detection S/N, can be significantly different when detrending is optimised using each of the two metrics, especially in the case of low S/N detections.

It may not always be possible to determine after the fact whether a detection has been falsely inflated by optimisation bias. Since the optimisation bias is intrinsic to the method, it is therefore necessary to examine the optimisation methods themselves, rather than the results produced, in order to evaluate the bias. We do this as follows.

\subsection{Measuring the Optimisation Bias} \label{measure_bias}

We now examine the bias induced in the detection S/N by each detrending optimisation. To do this, we optimise the detrending parameters and calculate an optimised S/N at each and every point in planetary velocity space. Considering each point across \kp - \vwind space individually, we find the number of PCA iterations to apply in detrending such that the derived S/N at that point, from each of \ccfobs and \deltaccfa, is maximised. The observed S/N corresponding to that number of PCA iterations is then found at each point. Excluding from consideration the central band around the real planetary signal ($\lvert V_{\mathrm{wind}} \rvert$ < 10 \kmsa), the distribution of S/N values obtained covers regions of velocity space devoid of the majority of this signal. As a result, the statistical expectation is that a robust optimisation method should yield S/N values normally distributed about zero. On the other hand, an optimisation method which is systemically inflating detection significances should produce a shifted S/N distribution with median $>$ 0.

\begin{figure}
     \centering
     \begin{subfigure}[h]{0.5\textwidth}
         \centering
         \includegraphics[width=\textwidth]{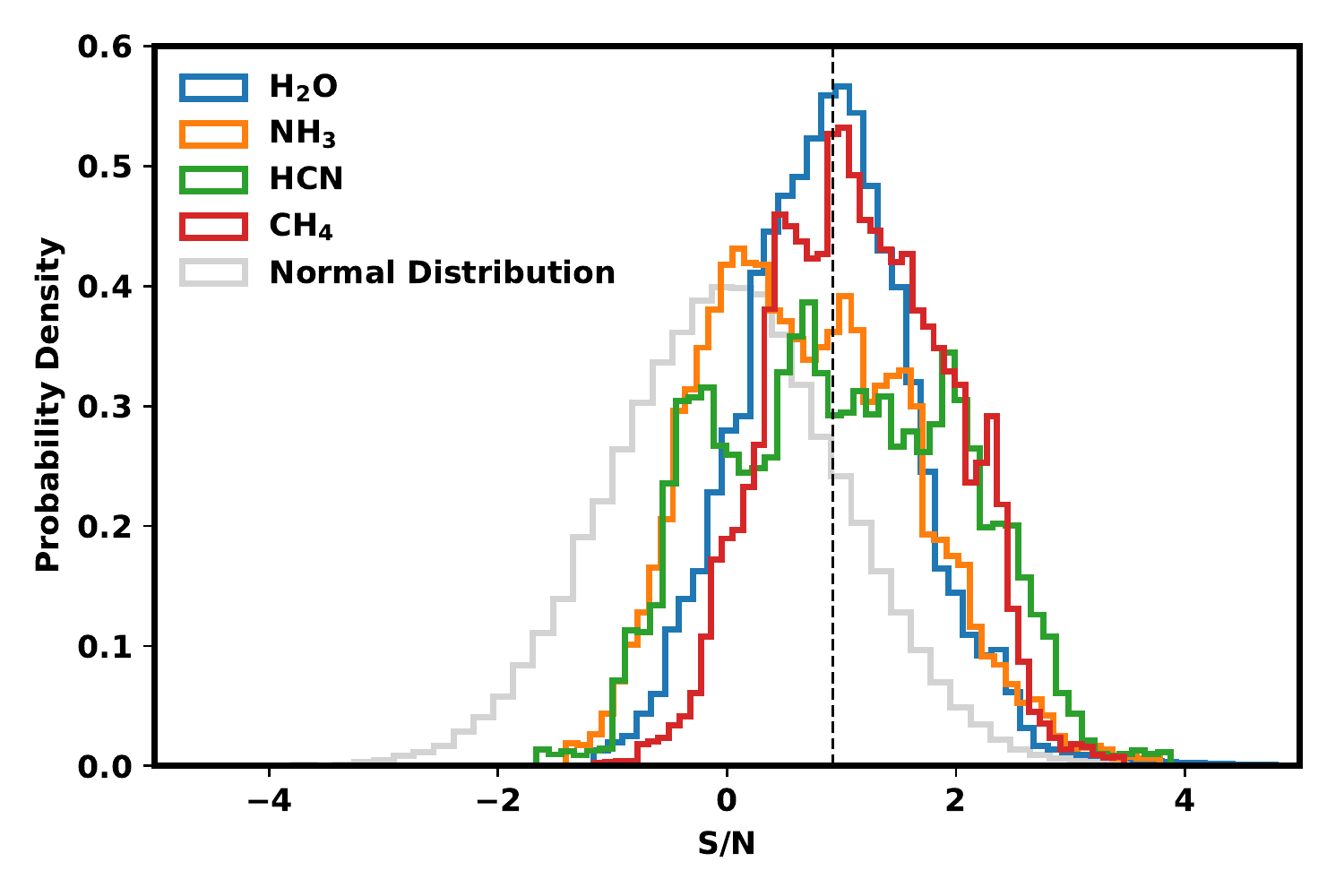}
         \caption{}
        \label{fig:hd189733b_data_sn}
     \end{subfigure}
     \hfill
     \begin{subfigure}[h]{0.5\textwidth}
         \centering
         \includegraphics[width=\textwidth]{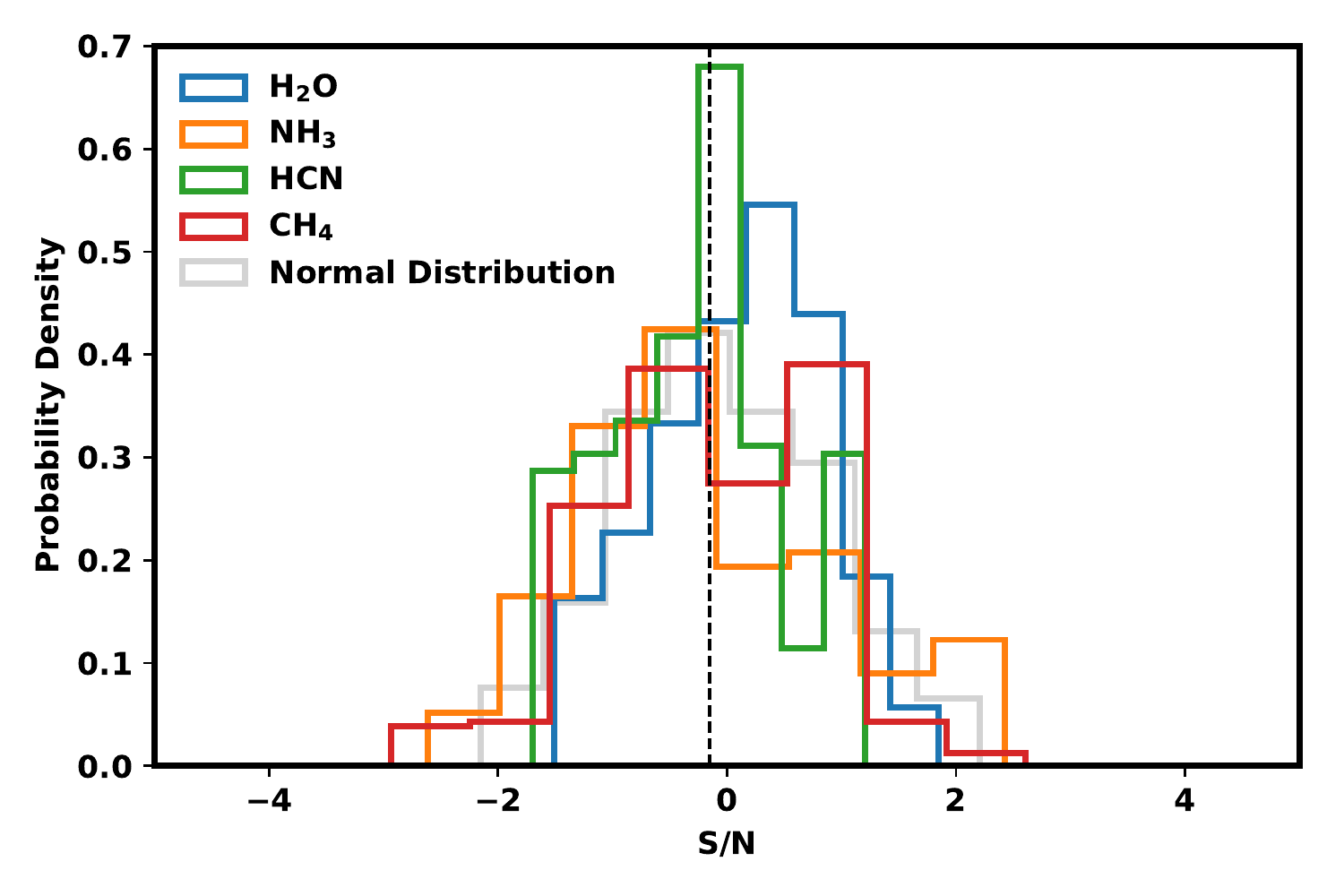}
         \caption{}
        \label{fig:hd189733b_delta_sn}
     \end{subfigure}
 \caption{The distribution of optimised S/N values across planetary velocity space when optimising everywhere the S/N from each of \ccfobs and \deltaccfa. In each case, the central band around the expected planetary velocity ($\lvert V_{\mathrm{wind}} \rvert$ < 10 \kmsa) is excluded from consideration, and the black vertical line represents the median of the combined distribution of all 4 models. An equally-populated normal distribution about zero is shown for reference in each panel. Panel (a): when the S/N from \ccfobs is optimised everywhere in the detrending, a distribution of S/N values with a median of 0.9 is found across the 4 models, demonstrating the inflated detection significances achieved using this method. Panel (b): when the S/N from \deltaccf is optimised everywhere in the detrending, a distribution of S/N values with a median of -0.2 is found across the 4 models, suggesting that this method is more robust.}
 \label{fig:sn_distribution}
\end{figure}

Figure \ref{fig:sn_distribution} shows the distribution of optimised S/N values across planetary velocity space, when optimising the S/N from each of \ccfobs and \deltaccf in the detrending. PCA iterations from 2 to 18 inclusive are considered, except for \ce{H2O} where a minimum of 3 iterations is enforced to aid the sufficient removal of telluric residuals. Across \ce{H2O}, \ce{NH3}, \ce{HCN} and \ce{CH4} models, the median optimised S/N across velocity space is 0.9 and -0.2 for \ccfobs and \deltaccfa, respectively. This suggests that optimising the detrending parameters using the S/N from \ccfobs biases the detection significance due to the method's vulnerability to noise. On the other hand, optimising the S/N from \deltaccf produces a smaller, negative bias. Whilst this is not zero, its magnitude is consistent with the median S/N values found away from the expected planetary velocity in Figure \ref{fig:hd189733b_h2o}, in which the optimisation of detrending was done only at the expected planetary velocity. Therefore, we do not observe an average increase in the S/N found at points away from the expected planetary velocity when the S/N from \deltaccf is optimised at every point in this space during detrending i.e. no bias is observed to be introduced by this detrending optimisation.

The bias could alternatively be measured by building a distribution of optimised S/N values at the expected planetary velocity for a large sample of randomised model spectra. Each random model is a \ce{H2O} model whose transit depth values have been randomly scrambled in wavelength space. There should therefore be no signal present in the CCF, and hence a detection S/N of zero is expected. For each random model, the detrending is optimised at the expected planetary velocity and a S/N is calculated. As before, the median of the S/N distribution returned can provide a measure of the optimisation bias induced at the expected planetary velocity.

A similar method of estimating optimisation bias at the expected planetary velocity is to randomise the time-ordering of spectra, rather than the wavelength-ordering of the model spectrum, prior to detrending \citep{zhang_platon_2020, giacobbe_five_2021}. This results in the planetary signal no longer being sinusoidally Doppler-shifted with time, and therefore no peak in velocity space should be recovered. However, due to the short transit duration of hot Jupiters, and the therefore narrow range in planetary radial velocity during the observations, correlated signals may remain in the randomly ordered spectra \citep{giacobbe_five_2021}. Since significant peaks could hence be recovered even if the detrending is robust, we do not use this method to measure the optimisation bias.

\begin{figure}
    \centering
    \includegraphics[width=0.5\textwidth]{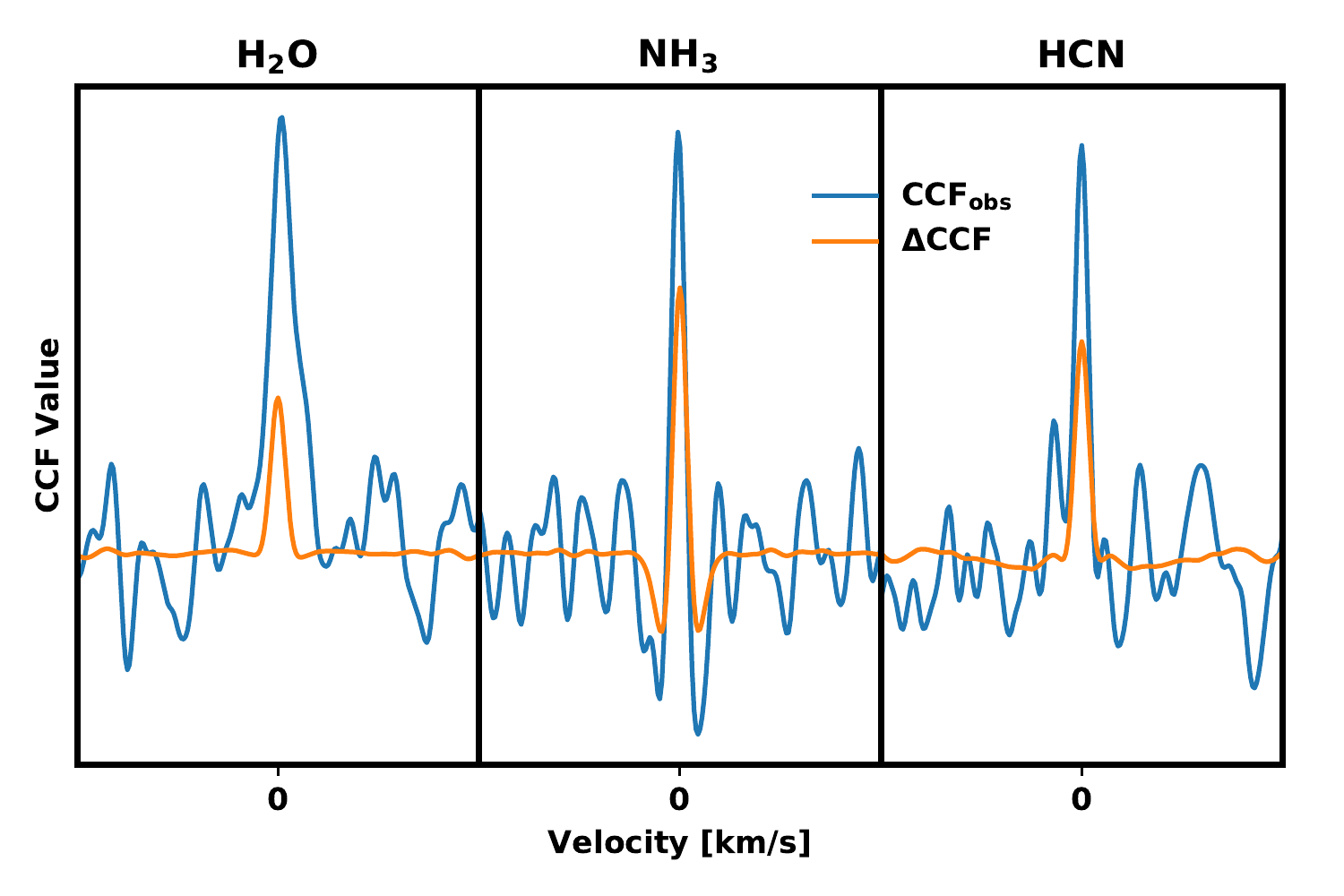} 
    \caption{A comparison between the optimised total \ccfobs (blue) and the total \deltaccf (orange) for a model signal injected into an arbitrary point in planetary velocity space away from the expected planetary velocity. The S/N from \ccfobs is optimised in the detrending, with both signals shown after the same number of PCA iterations. The total \deltaccf is the isolated contribution of the injected signal to \ccfobsa. As can be seen, the peak of the optimised total \ccfobs is greater than the peak of the injection itself, directly demonstrating the amplification of noise into a more significant detection. The planetary velocity considered for each molecule is not necessarily the same, nor is the absolute scale of the CCF value axis.}
    \label{fig:hd189733b_bias_plot}
\end{figure}

We now provide an illustration of how optimising the S/N from \ccfobs in the detrending can amplify noise into a more significant detection. We inject a model signal into an arbitrary point in velocity space away from the real signal. This injected signal is now treated as a real signal in the data. To recover this signal, during detrending we optimise the S/N from \ccfobs (which contains the injected signal) at the injected velocity. Figure \ref{fig:hd189733b_bias_plot} compares the optimised total \ccfobs to the total \deltaccfa, which is the isolated contribution of the injected signal to \ccfobsa, after the same number of PCA iterations. The optimised \ccfobs signal is greater than the injected signal itself due to the amplification of noise resulting in a more significant detection. When optimising the S/N from \ccfobsa, such systematic amplification leads to the bias observed in Figure \ref{fig:sn_distribution}.

We conclude that bias is introduced when the S/N from \ccfobs is optimised at the expected planetary velocity in the detrending, whereas using \deltaccf is more robust.

\subsection{Detrending Performance} \label{performance}

When judging the robustness of a detrending optimisation method, it is important to consider how a good detrending method should perform. An ideal detrending method would remove telluric, stellar and instrumental effects from our spectra, leaving only the planetary signal and white noise. Ideally, such removal would be independent of the velocity shift (assuming a sufficiently large change in the planet's radial velocity over the transit), strength and model of the signal for which we are optimising. It would also not matter if this signal is actually present in the data or not. The optimal detrending parameters derived would therefore show little variation across velocity space and between different models. In light of these expectations, we here investigate the detrending behaviour when optimising the S/N from each of \ccfobs and \deltaccfa.

\begin{figure*}
     \centering
     \begin{subfigure}[h]{0.33\linewidth}
         \centering
         \includegraphics[width=\textwidth]{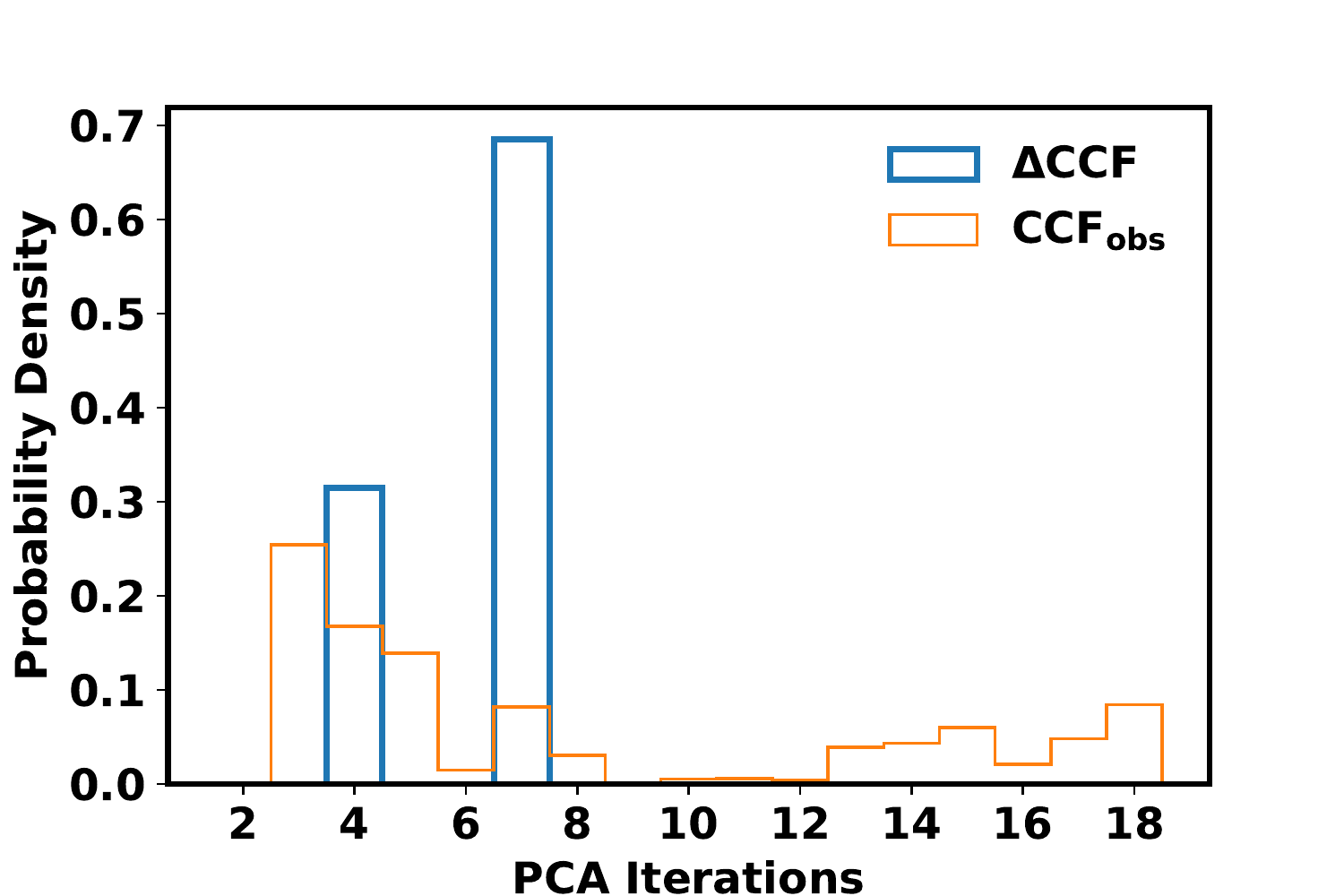}
         \caption{\ce{H2O}}
         \label{fig:hd189733b_h2o_detrending_hist}
     \end{subfigure}
     \begin{subfigure}[h]{0.33\linewidth}
        \centering
        \includegraphics[width=\textwidth]{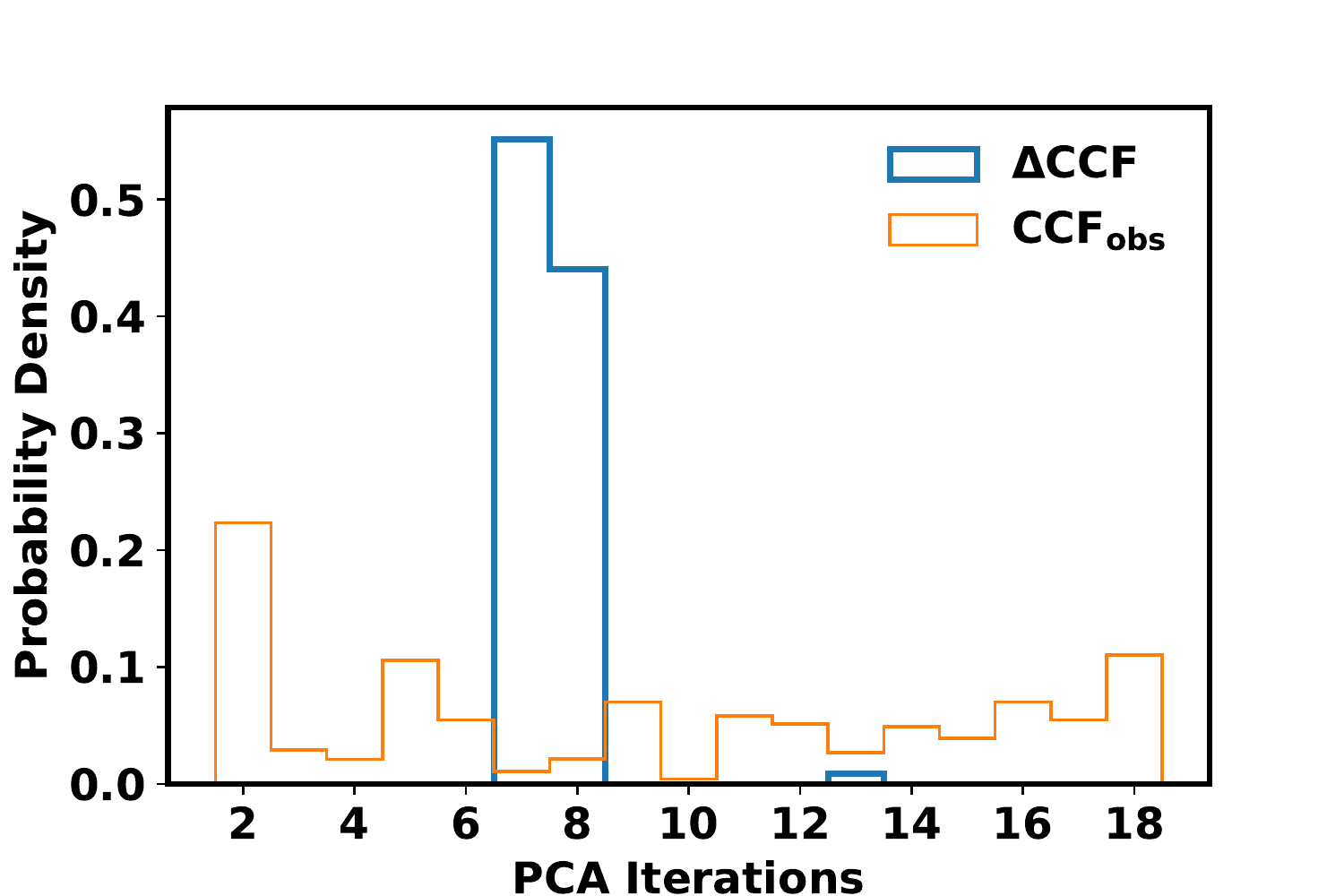}
        \caption{\ce{NH3}}
        \label{fig:hd189733b_nh3_detrending_hist}
     \end{subfigure}     
     \begin{subfigure}[h]{0.33\linewidth}
        \centering
        \includegraphics[width=\textwidth]{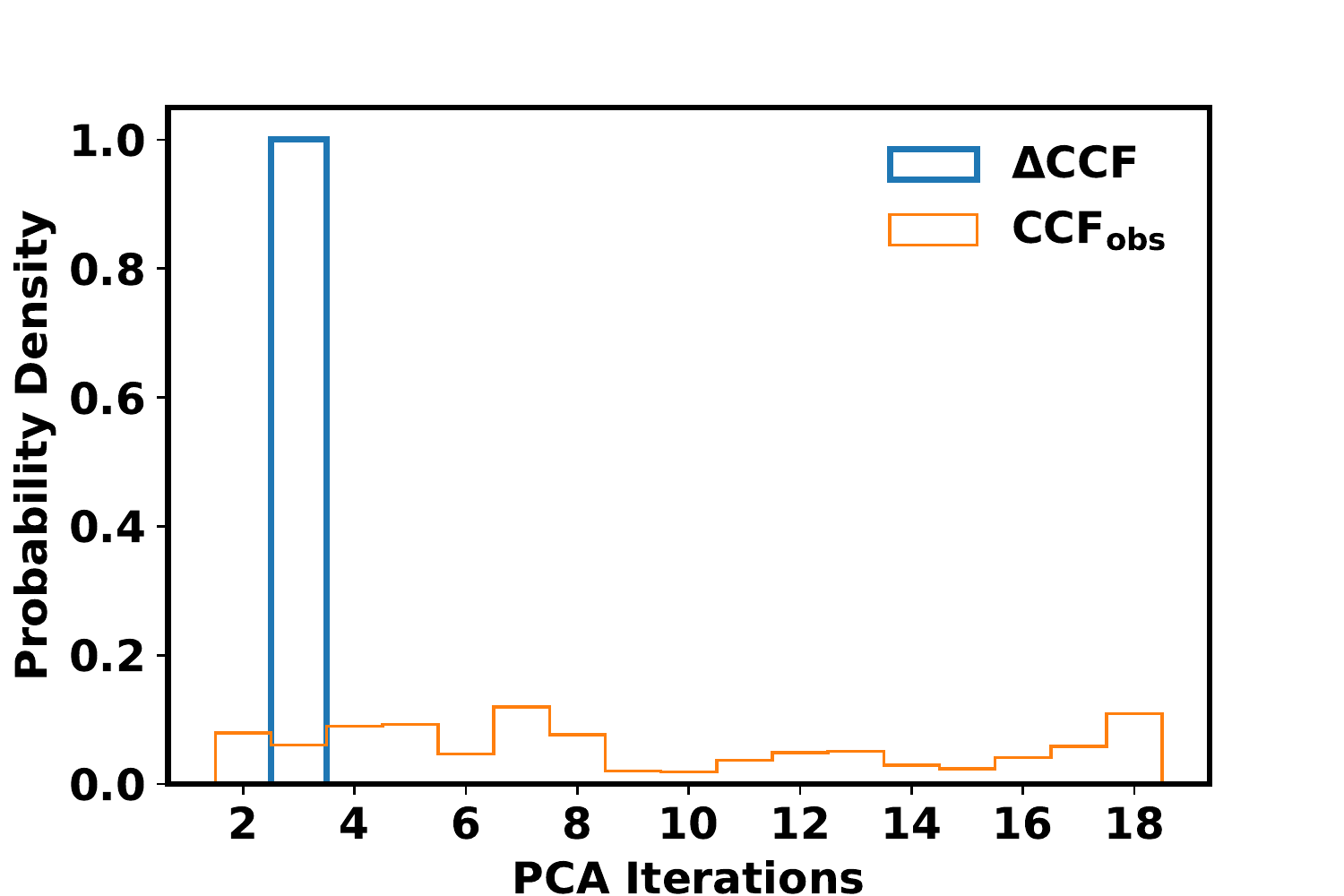}
        \caption{\ce{HCN}}
        \label{fig:hd189733b_hcn_detrending_hist}
     \end{subfigure}
 \caption{The distribution of the optimum PCA iteration across velocity space for each of \ce{H2O}, \ce{NH3} and \ce{HCN}. Optimising the S/N from \deltaccf (blue) gives a tightly constrained distribution for this parameter across velocity space. There is also consistency between models. On the other hand, detrending parameters derived by optimising the S/N from \ccfobs show large variation across planetary velocity space and between models.}
 \label{fig:detrending_histograms}
\end{figure*}

Figure \ref{fig:detrending_histograms} shows the optimal detrending parameters derived by optimising the S/N from each of \ccfobs and \deltaccf over velocity space and between different models. As in Section \ref{measure_bias}, we inject a model at each location in planetary velocity space and optimise its recovery using each of these metrics. Whereas Figure \ref{fig:sn_distribution} shows the optimised S/N distributions, in Figure \ref{fig:detrending_histograms} we show the distributions of the corresponding detrending parameters used to optimise each of the metrics at each point in planetary velocity space. The relative consistency of optimising the S/N from \deltaccf across velocity space and between different models is demonstrated. For this dataset, across \ce{H2O}, \ce{NH3} and \ce{HCN} models, the S/N from \deltaccf is most commonly optimised by either 7-8 PCA iterations or 3-4 PCA iterations. This tightly constrained bimodal distribution is somewhat characteristic of the trend seen in Figure \ref{fig:hd189733b_sn_pca}, in which two local peaks in S/N appear at $\sim$4 and $\sim$7 PCA iterations for an \ce{H2O} model. These results therefore support the finding in Figure \ref{fig:hd189733b_sn_pca} that there is little change in the shape, and hence the optimum, of the \deltaccf S/N variation as we move across planetary velocity space, and the findings of Figure \ref{fig:signal_erosion} that the erosion of an injected planetary signal with each PCA iteration is somewhat consistent across models. No such consistency across velocity space is observed when optimising the S/N from \ccfobsa. The optimised detrending parameters in this case are highly dependent on planetary velocity, which is expected due to the noise in \ccfobs being variable across planetary velocity space.

Since the detrending parameters derived by optimising the S/N from \deltaccf appear consistent across velocity space and between the models we consider, as demonstrated in Figure \ref{fig:detrending_histograms}, it is unlikely that significant bias will be introduced due to the specific choice of injection velocity or atmospheric model used in the detrending optimisation. For this dataset, we find that the set of likely optimised detrending parameters is relatively independent of such choices. We note, however, that we have not investigated more extreme models, e.g. \ce{CO2}-dominated atmospheres, considering that HD 189733 b is a gas giant with an H$_2$-rich atmosphere.

As discussed in Section \ref{delta_method}, noise around the injected planetary velocity is subtracted and therefore not considered when calculating the S/N from \deltaccfa. The robustness of the derived detrending parameters against velocity, as demonstrated here, may however suggest that such loss of noise is perhaps not overly consequential. At each different planetary velocity, different regions of the CCF are not considered, but similar optimised detrending parameters are found.

\subsection{Extension to Order-wise Optimisation} \label{order-wise}

Given the varying telluric contamination and planetary signal strength in each spectral order, it is reasonable to assume that a different amount of detrending is required for each order \citep{alonso-floriano_multiple_2019, spring_black_2022}. However, order-wise optimisation is typically avoided due to the significantly greater number of free parameters in the analysis, which increases the risk of amplifying noise into false detections \citep{cabot_robustness_2019, spring_black_2022}. We investigate this by optimising separately the number of PCA iterations applied to each order. Again, we consider iterations between 3 and 18 for \ce{H2O}, and between 2 and 18 for other species. 

\begin{figure*}
     \centering
     \begin{subfigure}[h]{0.49\textwidth}
         \centering
         \includegraphics[width=\textwidth]{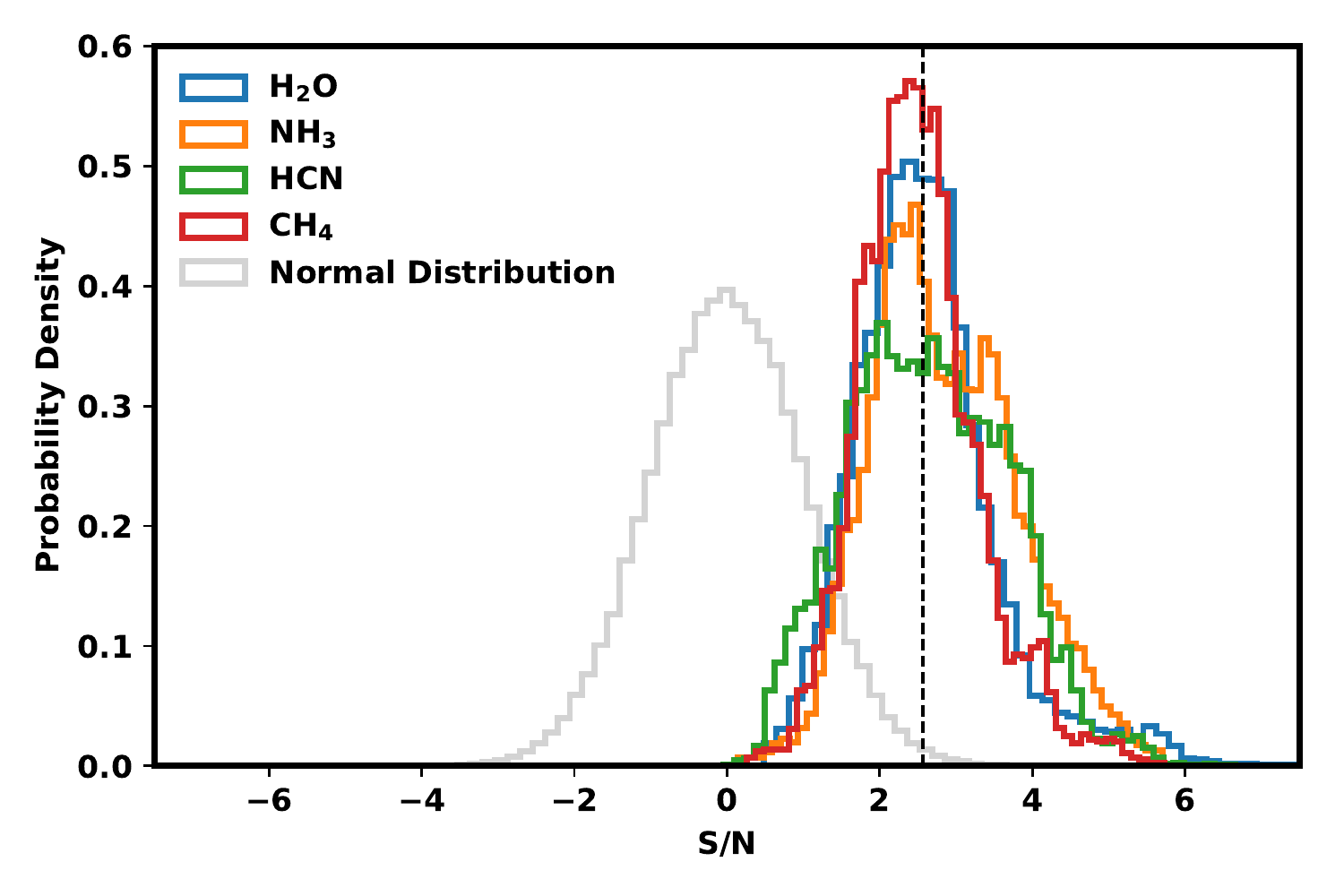}
         \caption{}
        \label{fig:hd189733b_data_orderwise_sn}
     \end{subfigure}
     \hfill
     \begin{subfigure}[h]{0.49\textwidth}
         \centering
         \includegraphics[width=\textwidth]{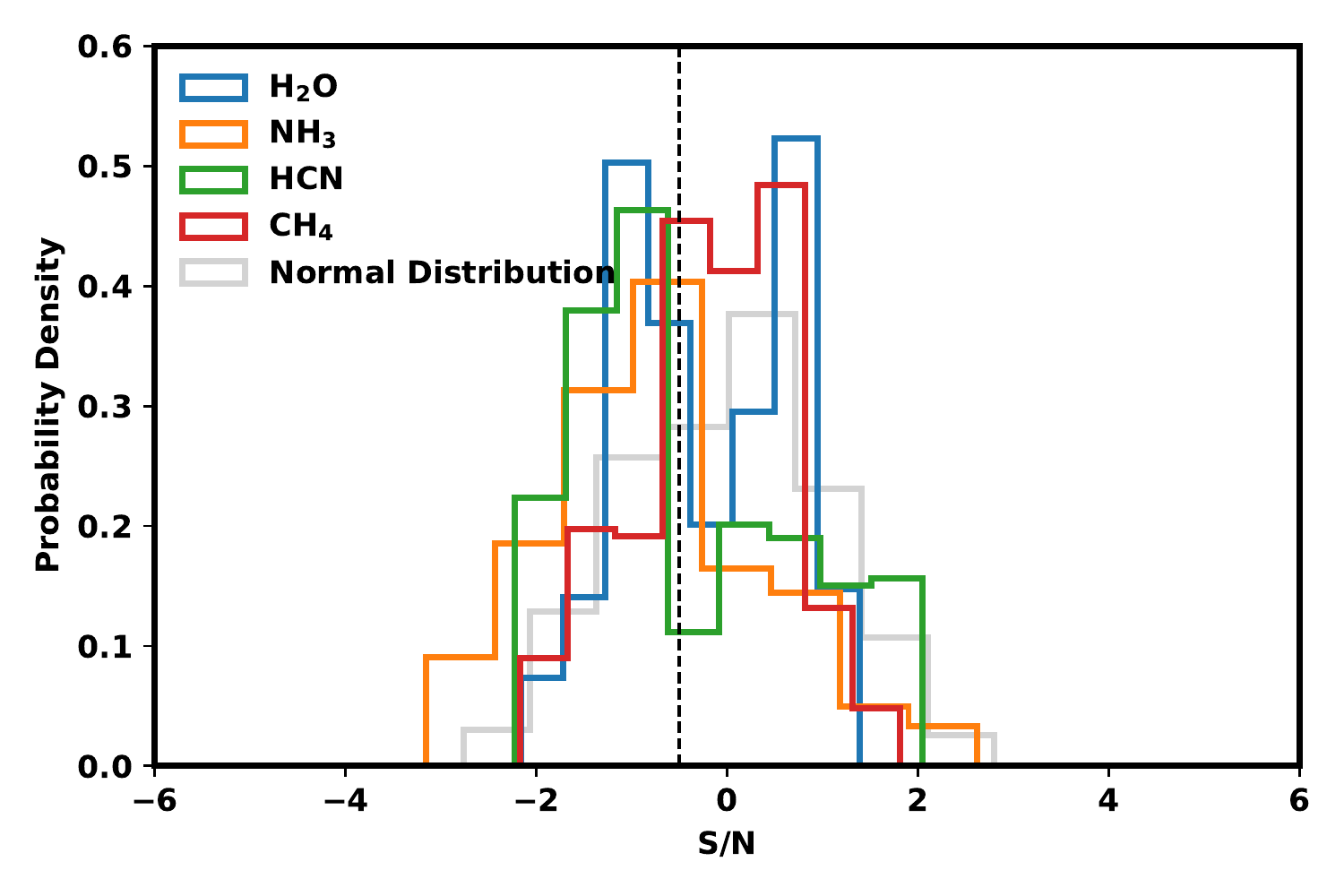}
         \caption{}
        \label{fig:hd189733b_delta_orderwise_sn}
     \end{subfigure}
     \begin{subfigure}[h]{0.49\textwidth}
        \centering
        \includegraphics[width=\textwidth]{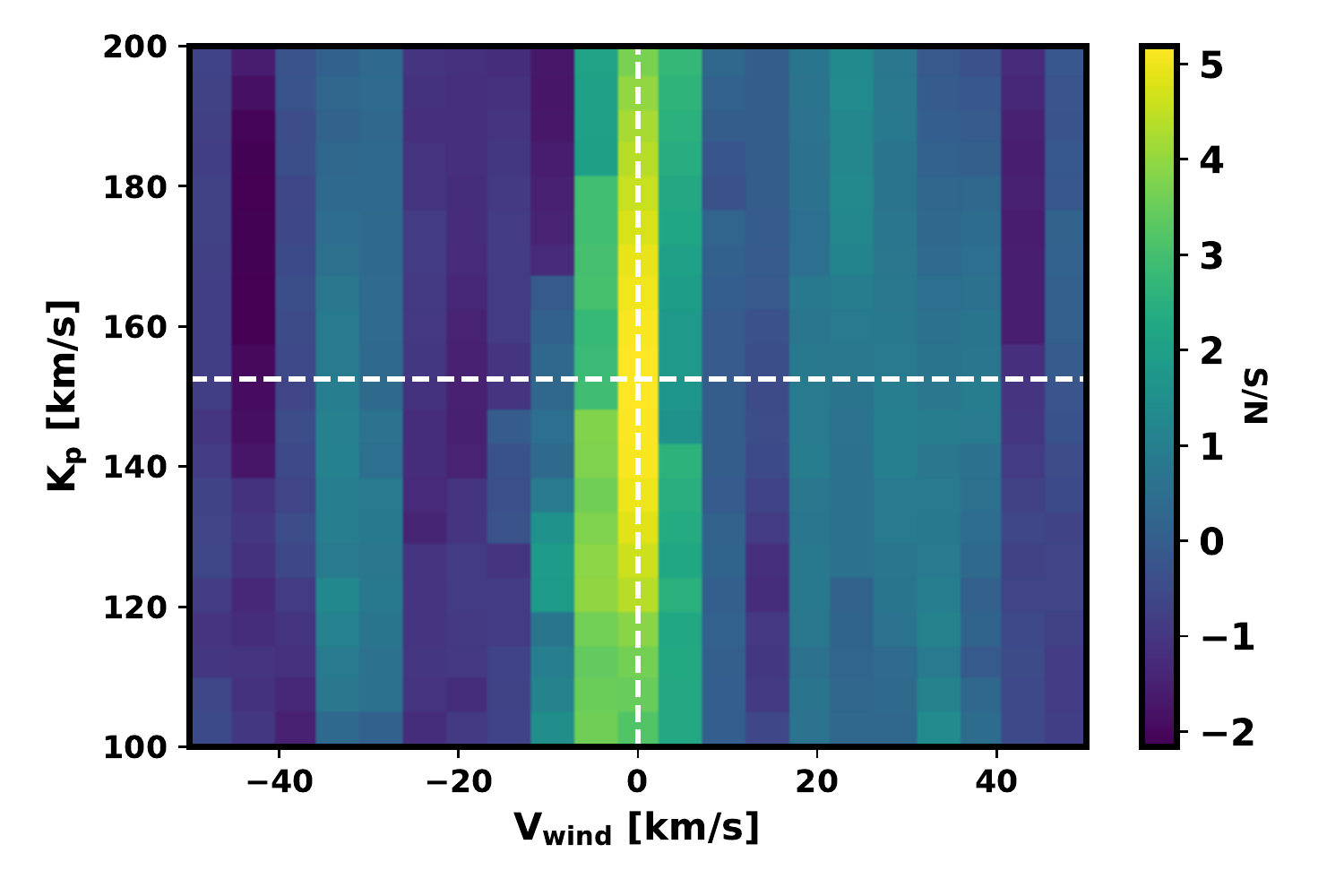}
        \caption{}
        \label{fig:hd189733b_h2o_delta_orderwise_everywhere}
     \end{subfigure}
 \caption{The S/N is optimised order-wise at every point across planetary velocity space. This is done by optimising order-wise the S/N from each of \ccfobs and \deltaccf in the detrending. The distributions of optimised S/N values across planetary velocity space are shown. In panels (a) and (b), the central band around the expected planetary velocity ($\lvert V_{\mathrm{wind}} \rvert$ < 10\kmsa) is excluded from consideration, and the black vertical lines represent the median of the combined distributions of all 4 models. An equally-populated normal distribution about zero is shown for reference in each of these panels. Panel (a): when the S/N from \ccfobs is optimised order-wise in the detrending, a median S/N of 2.6 is found across the 4 models, demonstrating the bias introduced by this method. Panel (b): when the S/N from \deltaccf is optimised order-wise in the detrending, a median S/N of -0.5 is found across the 4 models, suggesting that this method is more robust. Panel (c): for \ce{H2O}, the optimised S/N values are now shown as a function of velocity space, when the S/N from \deltaccf is optimised order-wise in the detrending. The distribution of these S/N values (for $\lvert V_{\mathrm{wind}} \rvert$ > 10 \kmsa) is shown in panel (b).}
 \label{fig:sn_distribution_orderwise}
\end{figure*}

To assess the robustness of order-wise detrending, we again optimise the S/N for each and every point in planetary velocity space, this time allowing different numbers of PCA iterations to be applied to each order. We do this optimisation using the S/N from each of \ccfobs and \deltaccfa. The distributions of S/N values retrieved away from the expected planetary velocity are shown in Figure \ref{fig:sn_distribution_orderwise}. When the S/N from \ccfobs is optimised order-wise in the detrending (Figure \ref{fig:hd189733b_data_orderwise_sn}), a median S/N of 2.6 is found across the 4 models, demonstrating that a large bias is present when using this method. This bias is considerably greater than in the case of global detrending. We find that 32$\%$ and 9$\%$ of points in velocity space return S/N $\ge$3 and $\ge$4, respectively. This is in agreement with \cite{cabot_robustness_2019}, who showed that detection significances of more than 4$\sigma$ can be obtained by optimising the detrending order-wise at incorrect locations in planetary velocity space. These findings suggest that optimising the S/N from \ccfobs order-wise is vulnerable to the recovery of spurious signals with significant S/N values, or the inflation of weak signals into much stronger ones dominated by an amplified noise component.

\begin{figure*}
     \centering
     \begin{subfigure}[h]{0.49\linewidth}
         \centering
         \includegraphics[width=\textwidth]{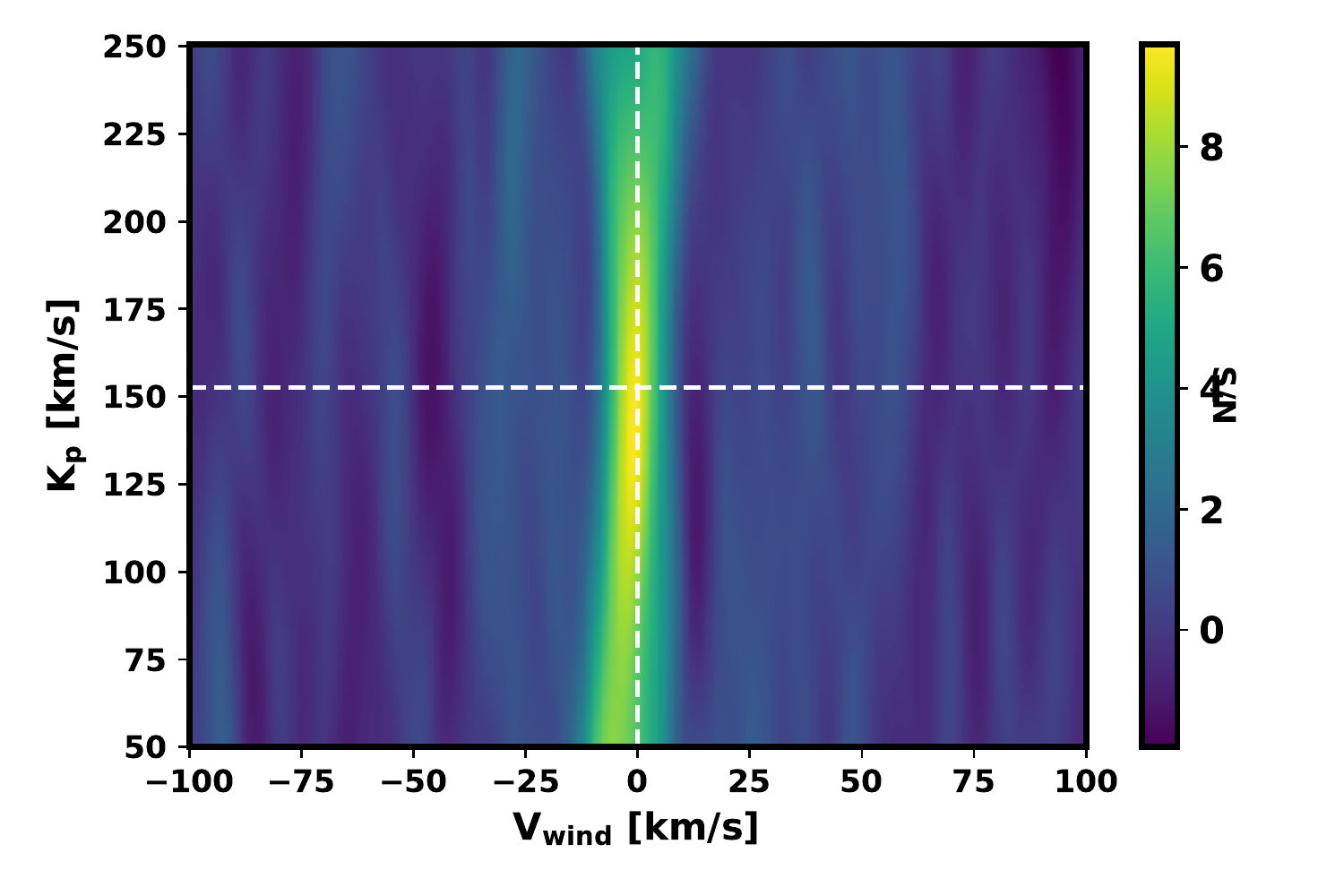}
         \caption{}
        \label{fig:189_h2o_data_orderwise}
     \end{subfigure}
     \hfill     
     \begin{subfigure}[h]{0.49\linewidth}
         \centering
         \includegraphics[width=\textwidth]{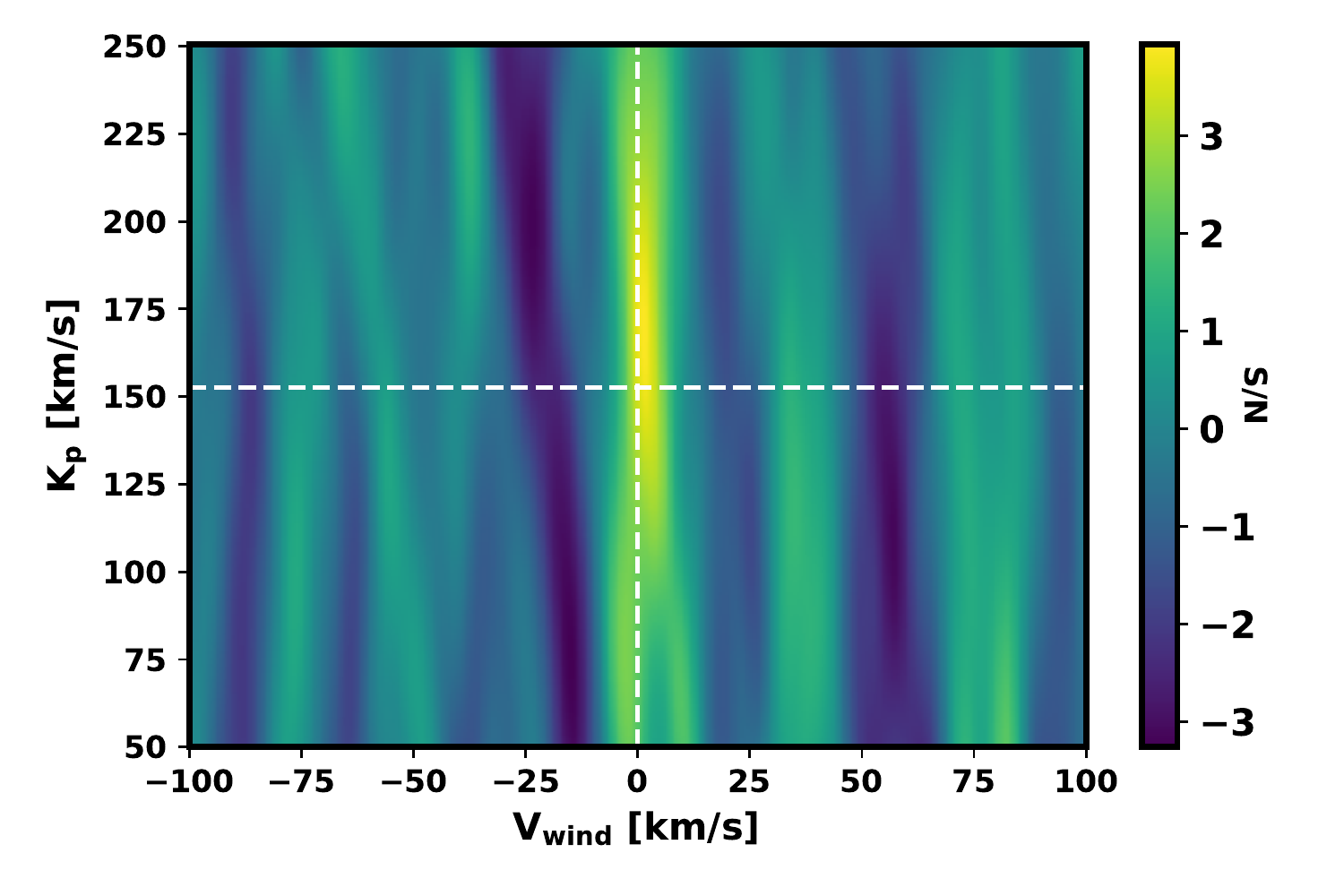}
         \caption{}
        \label{fig:189_nh3_data_orderwise}
     \end{subfigure}
 \caption{S/N maps showing non-robust signals for \ce{H2O} and \ce{NH3} in the atmosphere of HD 189733 b, via the order-wise detrending optimisation of the S/N from \ccfobsa. Panel (a): a S/N of 9.7 is retrieved for \ce{H2O}, with a median S/N across velocity space (for |\vwind| > 10 \kmsa) of 0.1. Panel (b): a S/N of 3.9 is retrieved for \ce{NH3}, with a median S/N across velocity space (for |\vwind| > 10 \kmsa) of -0.3.}
 \label{fig:189_data_orderwise}
\end{figure*}

We demonstrate the potential for these effects in Figure \ref{fig:189_data_orderwise}, in which we find significant signals for \ce{H2O} (S/N = 9.7) and \ce{NH3} (S/N = 3.9) in the atmosphere of HD 189733 b using order-wise optimisation of the S/N from \ccfobs during detrending. Using global detrending, or when optimising the S/N from \deltaccf order-wise, we are not able to recover a significant \ce{NH3} signal in this dataset. The signal may therefore be spurious, and only introduced by the bias attributed with optimising order-wise the S/N from \ccfobsa. Likewise, the significance of the \ce{H2O} signal is largely inflated compared to what was found robustly in Figure \ref{fig:hd189733b_h2o_7} (S/N = 6.1). There is no conclusive information in Figure \ref{fig:189_data_orderwise}, e.g. the median S/N across velocity space, which could indicate whether either detection S/N has been biased by contributions from residual noise. This motivates the above analysis of the optimisation methods and their intrinsic biases themselves, rather than just the results produced.

On the other hand, when the S/N from \deltaccf is optimised order-wise at each and every point across velocity space, a median S/N of -0.5 is found across the 4 models (Figure \ref{fig:hd189733b_delta_orderwise_sn}). Whilst this is again non-zero, there is no positive bias and the absolute bias is much less than that obtained using \ccfobsa. Figure \ref{fig:hd189733b_h2o_delta_orderwise_everywhere} shows the optimised S/N values across a region of velocity space for \ce{H2O}, with a clear and significant peak at the expected planetary velocity. We additionally find that, within each order, the distribution of optimised detrending parameters across planetary velocity space shows similar behaviour as in Figure \ref{fig:detrending_histograms}. We conclude that optimising the S/N from \deltaccf in each order is therefore more robust than using \ccfobsa.

\begin{figure}
    \centering
    \includegraphics[width=0.5\textwidth]{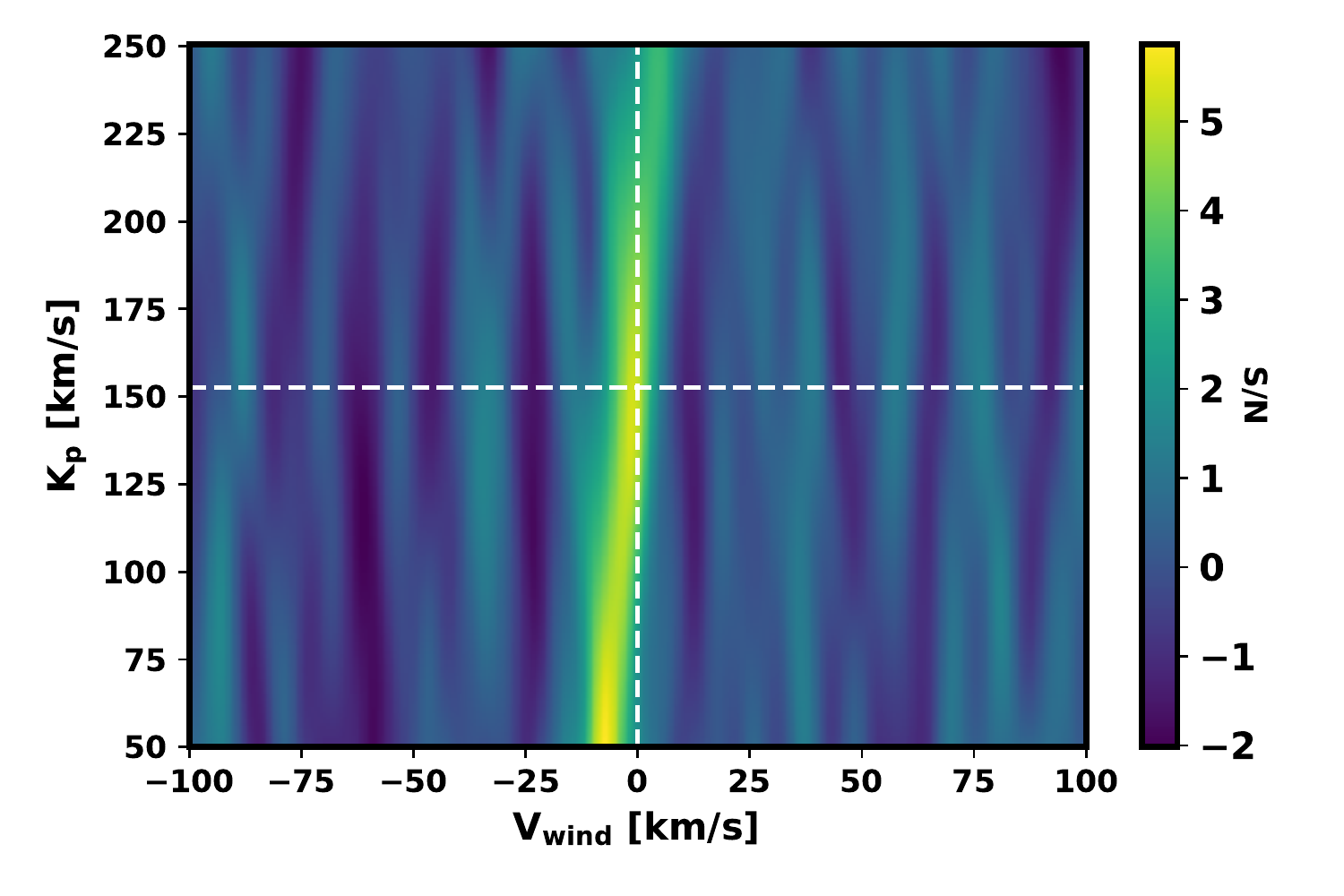} 
    \caption{S/N map showing the detection of \ce{H2O} in the atmosphere of HD 189733 b with a S/N of 5.4, via the order-wise detrending optimisation of the S/N from \deltaccfa.}
    \label{fig:189_h2o_delta_orderwise}
\end{figure}

Using this optimisation method, the retrieved S/N for \ce{H2O} is 5.4 (Figure \ref{fig:189_h2o_delta_orderwise}), which is consistent with that found via global detrending, albeit slightly lower. In other datasets and/or with different models, however, there may be an increase in S/N by allowing each order to be detrended separately. In this example, we note the persistent telluric contamination in the form of a spurious second peak at low \kp and negative $V_{\mathrm{wind}}$; this is discussed further in Section \ref{summary}. Despite this, the planetary signal at the expected planetary velocity does not appear to have been over-optimised to an inflated S/N, like that in Figure \ref{fig:189_h2o_data_orderwise}, suggesting that minimal bias is introduced at the expected planetary velocity by this detrending optimisation.

We therefore find that, as in the case of global detrending, significant bias is introduced when the S/N from \ccfobs is optimised order-wise at the expected planetary velocity in the detrending, whereas using \deltaccf is more robust.

\subsection{A Robust Detrending Recipe} \label{recipe}

We find that the selection of detrending parameters by optimising the S/N from \deltaccf will minimally bias the S/N at the expected planetary velocity, whether done globally or order-wise. We here summarise this method for obtaining a robust S/N measurement, starting with cleaned and normalised spectra:

\begin{enumerate}
    \item Apply iterations of PCA to the spectra. After each iteration, cross-correlate the residual with a model template (as described in Section \ref{signal_extraction}) to derive an observed cross-correlation function, \ccfobsa, for each iteration.
    
    \item Inject a model signal into the spectra at or close to the expected planetary velocity of the real signal. Repeat step (i) on the signal-injected spectra to derive a signal-injected cross-correlation function, \ccfinja, for each iteration. The planetary velocity of the injection can be somewhat approximate, as discussed in Section \ref{performance}.
    
    \item For each PCA iteration, derive the differential cross-correlation function, $\mathrm{\Delta CCF  =  CCF_{inj} - CCF_{obs}}$. Calculate the S/N from \deltaccf at the injected planetary velocity, as described in Sections \ref{signal_extraction} and \ref{delta_method}, and find the number of PCA iterations which maximises this S/N.
    
    \item Apply this optimal number of PCA iterations to the observed spectra. Cross-correlate the residuals with a Doppler-shifted model template, as in Section \ref{signal_extraction}, over planetary velocity space to derive the detection S/N as a function of planetary velocity.
\end{enumerate}

The above procedure is for global detrending optimisation. We have shown in Section \ref{order-wise} that it may also be robust to optimise separately the number of PCA iterations applied to each spectral order. In this case, steps 1-3 above can be applied to a single order to find the optimum number of PCA iterations required to detrend that order. The detrended residuals are then cross-correlated with the model template as before.

The choice of the maximum number of PCA iterations to consider during optimisation is somewhat arbitrary in the literature. In Figure \ref{fig:hd189733b_sn_pca}, the S/N from \deltaccf is clearly decreasing after 18 PCA iterations, hence this is a reasonable point at which to stop. This will not necessarily always be the case as we consider different datasets however. For a consistent determination of the maximum iteration to consider, we simulate the erosion of a model planetary signal, as in Figure \ref{fig:signal_erosion}. A minimum acceptable fraction of the remaining signal can be nominally defined, such that only iterations up to and including this are allowed. We here use $\sim$20-30$\%$ as this minimum fraction, and take the same maximum PCA iteration for different species due to the demonstrated consistency between the erosion of different models. We hence obtain a maximum of 18 PCA iterations for the HD 189733 b dataset.

It should be noted that, even when using identical detrending methods on the same dataset with the same model, the final S/N value will not necessarily be the same. We find that the addition of small amounts of Gaussian noise into the normalised spectra before detrending can produce a wide spread of observed S/N values. Unsubstantial differences in the cleaning, normalisation, calibration and masking of spectra prior to detrending can therefore considerably alter the reported detection significance, even when subsequent methods are identical.

\section{Additional Factors in Determination of Detection Significance} \label{factors}

In addition to the optimisation of detrending, there are other method and parameter choices within the data analysis which are inconsistent across the literature. It is important to understand the extent to which such choices can impact the detection S/N. In this section we present some examples.

\subsection{Velocity Range} \label{vel_range}

We now demonstrate how the retrieved detection S/N can be dependent on the planet-frame velocity range over which we calculate the noise in the total CCF. This dependency has previously been noted by \cite{spring_black_2022}. We use the globally detrended \ce{H2O} signal in the atmosphere of HD 189733 b to demonstrate this. Until now, we have cross-correlated the detrended spectra with the model template for velocities ranging from -400 \kms to 400 \kms in intervals of 1 \kmsa. The CCF is then shifted into the planet-frame, and the noise is calculated by taking the standard deviation of the total CCF values between velocities of -300 \kms to 300 \kmsa, excluding the $\pm$15 \kms region as described in Section \ref{signal_extraction}. We examine the dependence of the retrieved S/N on this velocity range over which we calculate the noise.

\begin{figure}
    \centering
    \includegraphics[width=0.5\textwidth]{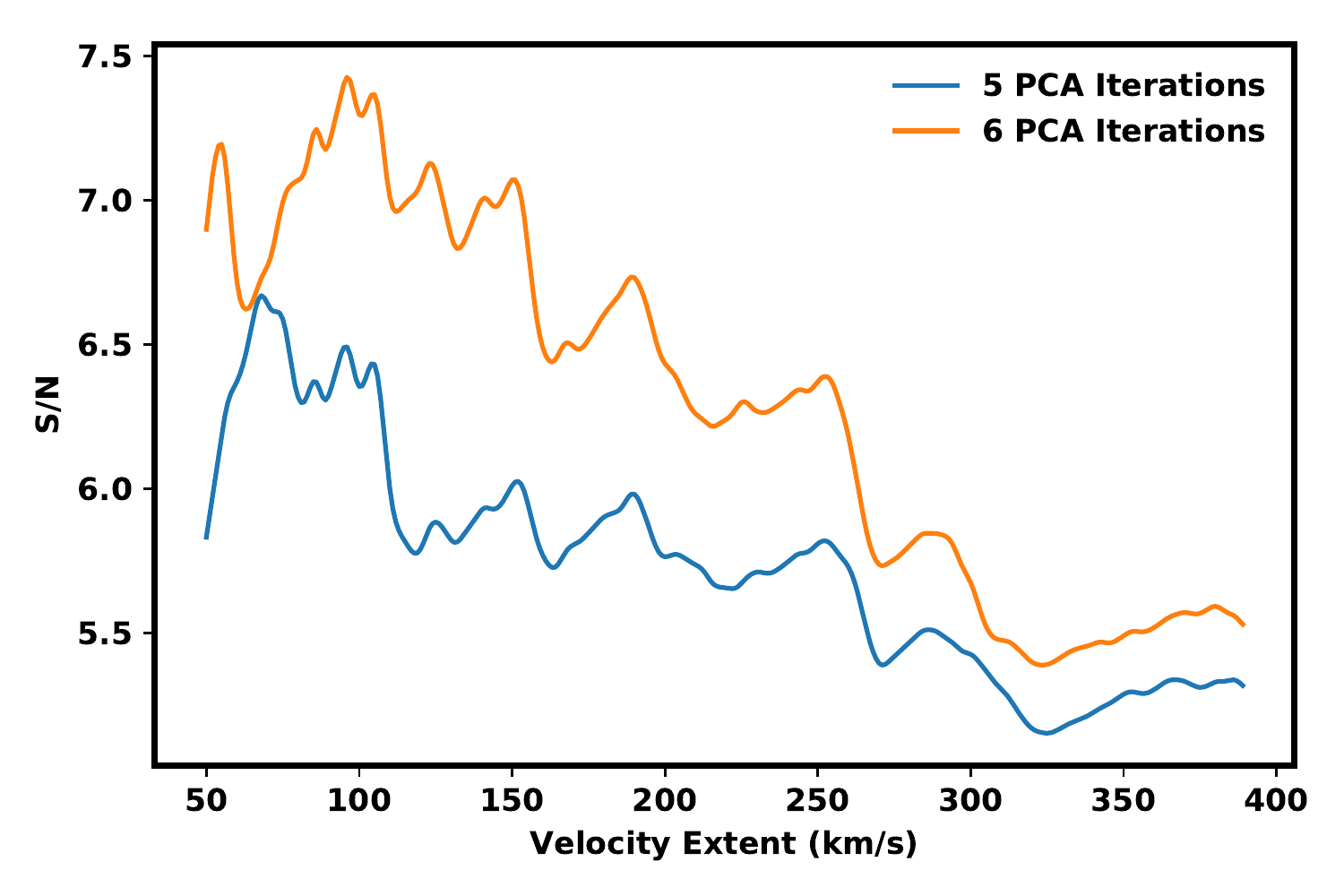}
    \caption{For \ce{H2O}, the retrieved S/N at the expected planetary velocity varies with the maximum velocity we consider when calculating the standard deviation of the total CCF away from the peak. This is shown for the cases of global detrending using 5 (blue) and 6 (orange) PCA iterations.}
    \label{fig:velocity_grid}
\end{figure}

Figure \ref{fig:velocity_grid} shows the variation of the retrieved S/N, for two different PCA iterations, against the maximum velocity we consider when calculating the standard deviation of the total CCF away from the central peak. Considerably amplified significances can be achieved when velocity ranges narrower than $\pm$150 \kms are considered. Values found here are in line with the detection of \ce{H2O} (S/N = 6.6) reported by \cite{alonso-floriano_multiple_2019}, where a velocity range of $\pm$65 \kms was used. Our velocity range of $\pm$300 \kms gives a more conservative detection significance.

When calculating the standard deviation of the total CCF, we therefore encourage the consideration of a wide velocity range to avoid domination by any relatively noiseless regions which can lead to amplified detection significances. Future work could develop a significance metric which is not so dependent on this parameter. For example, the signal and noise could perhaps be estimated using the mean and standard error of the CCF values within a pixel-wide in-trail distribution. In the meantime, potential amplification of the quoted detection S/N by this effect should be considered.

\subsection{Optimising Order Weighting and Selection} \label{weighting}

When optimising the detrending parameters order-wise, the number of PCA iterations which maximises the retrieved S/N in each order is selected. During global detrending, we can similarly calculate the S/N in each order but instead use it to weight each order's contribution to the total CCF. Previous works have done this to improve the S/N of a detection, by favouring the orders where the signal can be recovered to a higher significance \citep{giacobbe_five_2021, spring_black_2022, van_sluijs_carbon_2022}. This may be the case in certain orders due to there being fewer telluric and stellar lines, or more planetary signal. However, it may also be due to there being a greater noise component in the spectrum in some orders. Favouring such orders, where amplified noise at the expected planetary velocity is falsely increasing the signal in that order, could bias the detection significance. We here investigate the robustness of such order weightings.

One method of weighting orders is to mask orders where there is little recoverable signal. For example, \cite{giacobbe_five_2021} mask orders in which the S/N recovered from \ccfinj at the expected planetary velocity is less than a threshold. As already discussed, this is model dependent and therefore difficult to reproduce. We therefore here examine the robustness of masking a fixed percentage of orders, using the order-wise S/N values from each of \ccfobs and \deltaccf at the expected planetary velocity.

To do so, each spectral order is first detrended using the number of PCA iterations found by globally optimising the S/N from \deltaccfa. For this dataset, in the case of \ce{H2O}, this means that 4 PCA iterations are applied, such that the unweighted case is equivalent to the detection shown in Figure \ref{fig:hd189733b_h2o_7}. We then apply the same robustness tests as in Figures \ref{fig:sn_distribution} and \ref{fig:sn_distribution_orderwise}. For each and every point in planetary velocity space individually, we calculate the S/N by only including in the CCF the 75$\%$ of orders which recover the greatest order-wise S/N, from each of \ccfobs and \deltaccfa, at that planetary velocity. The remaining 25$\%$ of orders are masked. We then observe the resulting distribution of S/N values. In the \ccfobs case, the distribution of retrieved S/N values has a median $>$ 1, suggesting the introduction of a bias when orders are selected according to the order-wise S/N from \ccfobsa. On the other hand, selecting orders according to the order-wise S/N from \deltaccf returns a distribution of S/N values with a median that is small in magnitude, suggesting that this method is considerably less biased. These findings remain true when the percentage of orders selected is varied.

\begin{figure}
    \centering
    \includegraphics[width=0.5\textwidth]{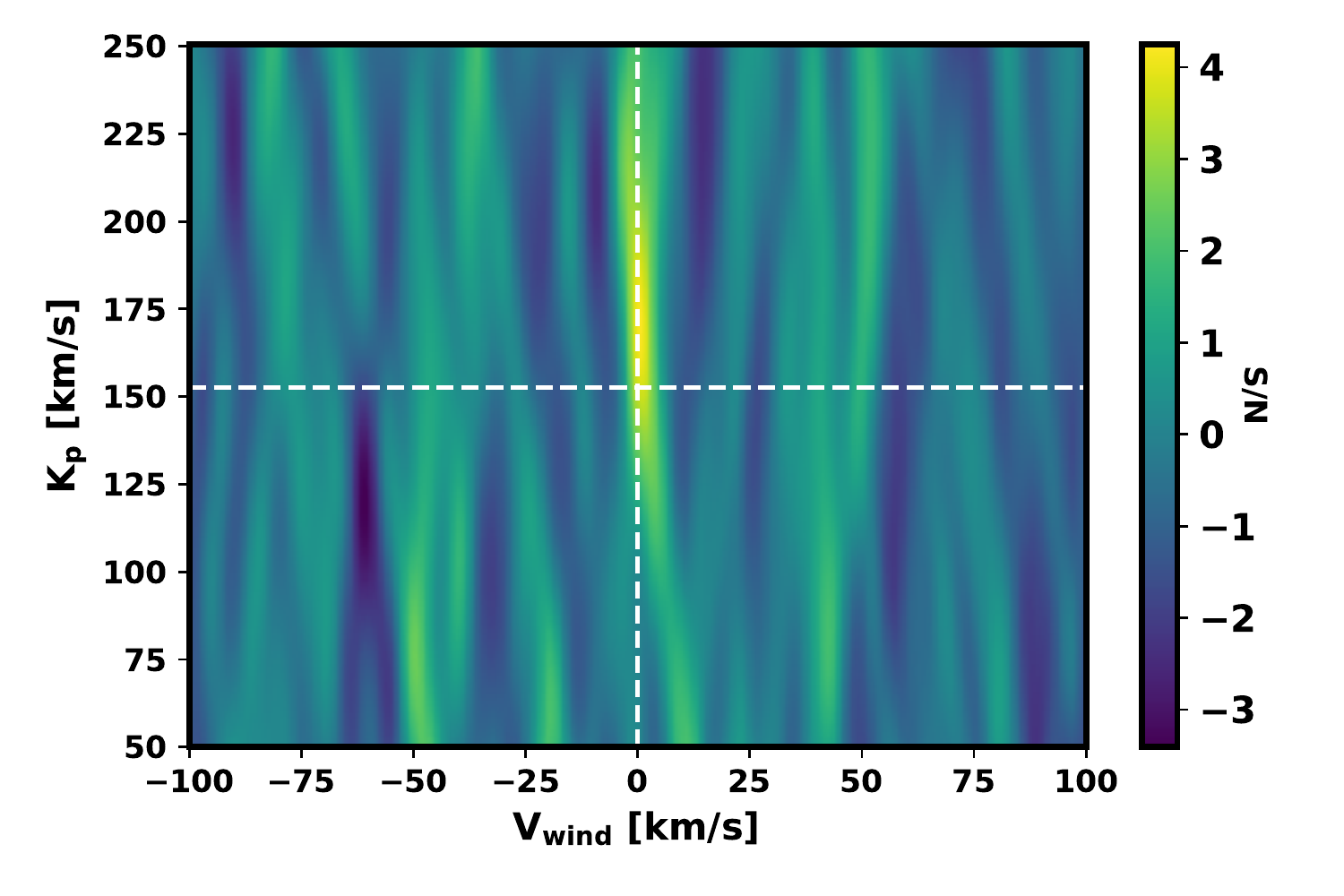}
    \caption{S/N map showing a non-robust \ce{NH3} signal (S/N = 4.2) in the atmosphere of HD 189733 b, achieved by masking orders according to the order-wise S/N from \ccfobsa. After global detrending using 17 PCA iterations, found by optimising the S/N from \ccfobs in the detrending, only the 15 out of 20 orders with the greatest S/N from \ccfobs at the expected planetary velocity are included in the final CCF.}
    \label{fig:hd189733b_189_nh3_order_selection}
\end{figure}

We explore examples to demonstrate these findings. As when optimising order-wise the S/N from \ccfobs during detrending, we can again retrieve an \ce{NH3} signal in the atmosphere of HD 189733 b, this time by selecting orders according to the order-wise S/N from \ccfobs at the expected planetary velocity. We first globally optimise the S/N from \ccfobs during detrending, corresponding to 17 PCA iterations, to find \ce{NH3} with a S/N of 3.2. We subsequently only select the 15 orders out of 20 with the greatest order-wise S/N from \ccfobs at the expected planetary velocity. An updated S/N of 4.2 is returned (Figure \ref{fig:hd189733b_189_nh3_order_selection}), providing a further example of the recovery of a tentative signal via non-robust methods.

\begin{figure}
    \centering
    \includegraphics[width=0.5\textwidth]{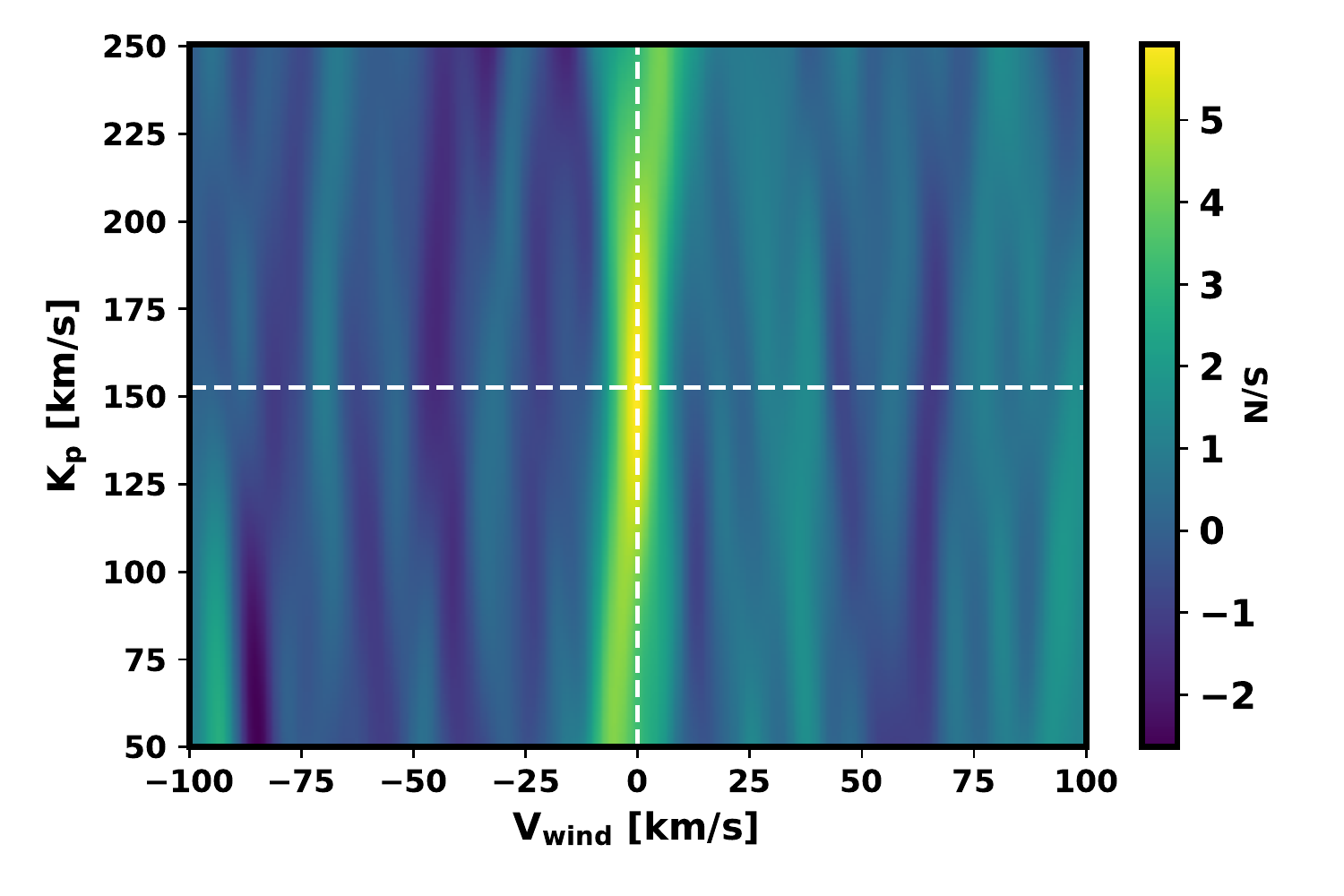}
    \caption{The HD 189733 b \ce{H2O} signal in Figure \ref{fig:hd189733b_h2o_7}, achieved via global detrending using 4 PCA iterations, is now robustly refined such that the final CCF only includes the 16 out of 20 orders with the greatest S/N from \deltaccfa at the expected planetary velocity. A new S/N of 5.9 is achieved, compared to 6.1 previously.}
    \label{fig:hd189733b_h2o_7_sel}
\end{figure}

We now instead apply robust order selection to our detection (S/N = 6.1) of \ce{H2O} in the atmosphere of HD 189733 b (Figure \ref{fig:hd189733b_h2o_7}), achieved via global detrending using 4 PCA iterations. We do so by considering the order-wise S/N from \deltaccf at the expected planetary velocity. There are 20 spectral orders remaining after a priori masking but only the 16 orders with the greatest S/N from \deltaccf are included when calculating the final CCF. An updated S/N of 5.9 is achieved (Figure \ref{fig:hd189733b_h2o_7_sel}), which is consistent with the original detection. In other datasets and/or with different models, it is possible that such robust order masking may result in an increased S/N.

Alternatively, unequal weightings, rather than a binary mask, can be applied to each order when their CCFs are summed \citep{spring_black_2022, van_sluijs_carbon_2022}. Thus far in this work the CCFs from each remaining order have been added with equal weighting. During this summation, we now instead weight the CCF from each order according to the order-wise S/N recovered at the expected planetary velocity. \cite{spring_black_2022} do so using the S/N from \deltaccf in each order. The weight \textit{w} for each \textit{i}-th order is calculated as:
\begin{equation} \label{weighting_eq}
\mathrm{w_{i} = \frac{S/N_{i} - S/N_{min}}{S/N_{max} - S/N_{min}}}
\end{equation}
where $\mathrm{S/N_{i}}$ is the order-wise S/N at the expected planetary velocity for order $i$, and $\mathrm{S/N_{min}}$ and $\mathrm{S/N_{max}}$ are the minimum and maximum of these S/N values across the orders, respectively.

We test this method for robustness as before. At each point in planetary velocity space, a final CCF is formed using order weightings calculated with equation \ref{weighting_eq}, using the order-wise S/N values from each of \ccfobs and \deltaccfa. A S/N value is then derived at each planetary velocity. In the \ccfobs case, the distribution of S/N values has a median $>$ 1, once again suggesting that a bias is introduced by this method. On the other hand, a median that is small in magnitude is obtained in the \deltaccf case, implying once more that this method is more robust. This follows the observed trend; optimising or weighting according to the S/N from \ccfobs is vulnerable to residual noise and therefore more biased, whereas \deltaccf is noise-subtracted and therefore its use is more robust. When these order weightings are applied to the globally detrended \ce{H2O} detection (S/N = 6.1) in Figure \ref{fig:hd189733b_h2o_7}, updated S/N values of 5.7 and 6.8 are found for the \deltaccf and \ccfobs cases, respectively.

To summarise, we find that a bias will likely be introduced if the contribution of each order to the final CCF is calculated according to the order-wise S/N values from \ccfobs at the expected planetary velocity. We conversely find that doing so using \deltaccf is more robust.

\section{Results: Case Studies} \label{case_studies} 

This work has so far used CARMENES observations of HD 189733 b as a test case to investigate the robustness of molecular detections using high-resolution transmission spectroscopy of exoplanets. We now systematically implement the robust methodologies described in Sections \ref{methods}, \ref{recipe}, and \ref{weighting} to archival CARMENES observations of two other hot Jupiters: HD 209458 b and WASP-76 b. Planet-specific models are independently generated for each dataset.

\subsection{HD 189733 b}

We here summarise our findings for the CARMENES observations of HD 189733 b. Following the robust methods described in Sections \ref{methods} and \ref{recipe}, we find \ce{H2O} in the atmosphere of HD 189733 b with a S/N of 6.1 (Figure \ref{fig:hd189733b_h2o_7}) when globally optimising the S/N from \deltaccf during detrending. S/N values of 5.4 and 5.9 are alternatively found when optimising the S/N from \deltaccf order-wise (Figure \ref{fig:189_h2o_delta_orderwise}), or when using robust order selection as detailed in Section \ref{weighting} (Figure \ref{fig:hd189733b_h2o_7_sel}), respectively. We do not robustly find any significant detections (S/N > 3.0) for \ce{NH3} or \ce{HCN} in the atmosphere of this exoplanet using this dataset.

We also outline our results when applying non-robust methods. By optimising the S/N from \ccfobs order-wise in the detrending, a S/N of 9.7 can be achieved for \ce{H2O} using the same spectra and model as before (Figure \ref{fig:189_h2o_data_orderwise}). This emphasises the extent to which detection significances can be inflated by non-robust detrending optimisations. We additionally obtain \ce{NH3} signals using non-robust methods, such as when optimising the S/N from \ccfobs order-wise in the detrending (S/N = 3.9) (Figure \ref{fig:189_nh3_data_orderwise}), and when selecting orders according to the order-wise S/N from \ccfobsa, after globally optimising the S/N from \ccfobs during detrending (S/N = 4.2) (Figure \ref{fig:hd189733b_189_nh3_order_selection}). Such detections of \ce{NH3} in the atmosphere of HD 189733 b are not necessarily spurious, but we can only obtain them here using non-robust methods.

\subsection{HD 209458 b} \label{hd209458b}

We now re-analyse archival CARMENES data for a transit of the hot Jupiter HD 209458 b on the night of 5th September 2018. HD 209458 b \citep{charbonneau_detection_2000} is another extensively studied hot Jupiter which has been subject to multiple high-resolution studies in the NIR region. \ce{H2O} has previously been detected using the same CARMENES observations as we use here \citep{sanchez-lopez_water_2019}, whilst observations from other high-resolution spectrographs have also found \ce{H2O} in both emission \citep{hawker_evidence_2018} and transmission \citep{giacobbe_five_2021}. \cite{hawker_evidence_2018} additionally detected \ce{CO} and \ce{HCN} in the atmosphere of HD 209458 b using two spectral bands of CRIRES (2.29–2.35 $\mu$m and 3.18–3.27 $\mu$m), whilst \cite{giacobbe_five_2021} found \ce{CO}, \ce{HCN}, \ce{CH4}, \ce{NH3} and \ce{C2H2} using GIANO over a spectral range of 0.95-2.45 $\mu$m. \ce{CO} was also detected by \cite{snellen_orbital_2010} in emission spectra observed using CRIRES.

\begin{figure}
    \centering
    \includegraphics[width=0.5\textwidth]{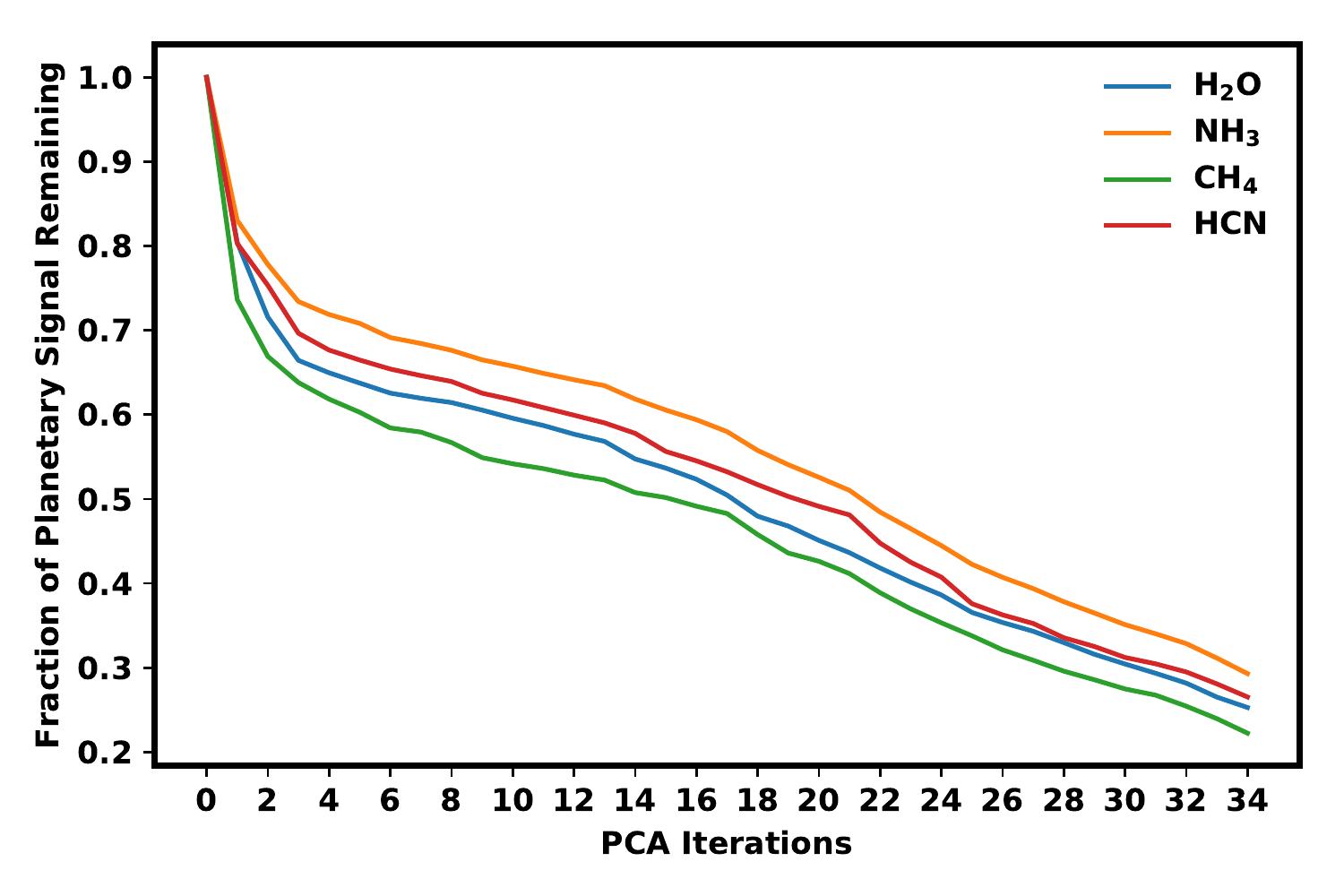}
    \caption{As in Figure \ref{fig:signal_erosion} but now for HD 209458 b, the total \deltaccf values of model planetary signals are eroded with each PCA iteration. \ce{H2O}, \ce{NH3}, \ce{CH4} and \ce{HCN} models are included here.}
    \label{fig:signal_erosion2}
\end{figure}

The data consists of 91 observations (spanning planetary orbital phases -0.0358 $< \phi <$ 0.0368), of which 46 are in transit. Exposure times of 198 s were used throughout. The median S/N of the observations was 76, with the airmass increasing from a minimum of 1.05 to a maximum of 2.11 over the observing night. The system parameters for this planet are shown in Table \ref{hd209458b_parameters}, and a planetary velocity semi-amplitude ($K_{\mathrm{p}}$) of $145\pm1.5$ \kms is expected \citep{giacobbe_five_2021}. We apply the robust methods presented in Sections \ref{methods} and \ref{recipe} to this dataset. We remove a priori the same 6 orders (54-53, 45-42) and similar exposures (first 15 and last 7 spectra) as \cite{sanchez-lopez_water_2019} due to low S/N. We find that the PCA erosion of the planetary signal is significantly slower for this dataset than that for HD 189733 b. As shown in Figure \ref{fig:signal_erosion2}, the nominal $\sim$20-30$\%$ threshold discussed in Section \ref{recipe} is reached only after roughly 34 iterations. We therefore consider PCA iterations up to and including 34 in our optimisations.

\begin{figure}
    \centering
    \includegraphics[width=0.5\textwidth]{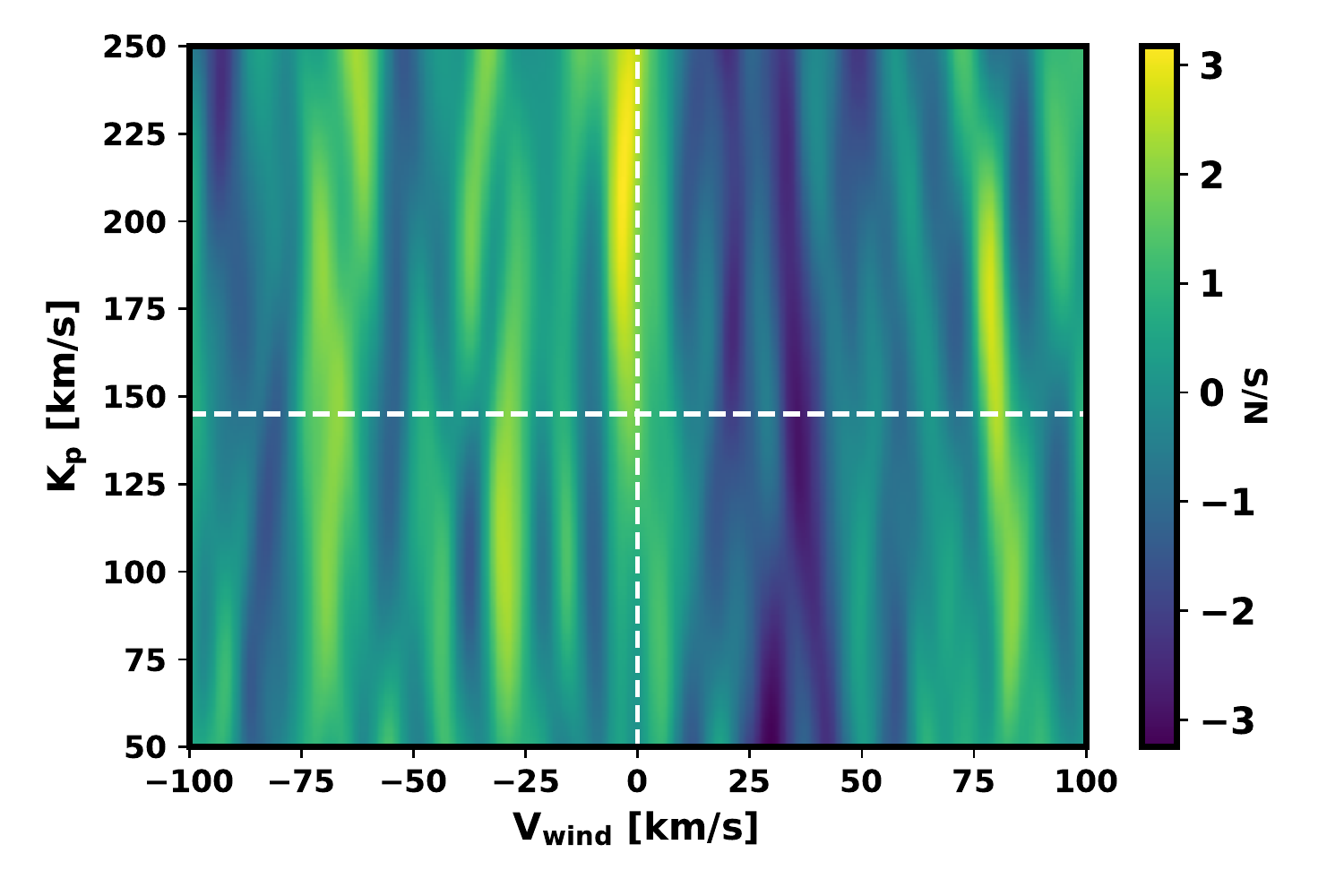}
    \caption{S/N map showing a tentative \ce{H2O} signal in the atmosphere of HD 209458 b, with a S/N of 3.2 and a significantly greater than expected value for $K_{\mathrm{p}}$. The S/N from \deltaccf was optimised in the detrending such that 21 PCA iterations were applied.}
    \label{fig:hd209458b_h2o_delta}
\end{figure}

We find a tentative signal for \ce{H2O}, with a S/N of 3.2 and a significantly greater than expected value for \kp (Figure \ref{fig:hd209458b_h2o_delta}), when globally optimising the S/N from \deltaccf during detrending. This corresponds to 21 PCA iterations. We are unable to robustly recover any signals for \ce{HCN}, which may be due to there being no strong spectral features in this wavelength band. \ce{NH3} and \ce{CH4} signals are also not observed. It is possible that a more comprehensive exploration of model space could yield stronger signals using robust methods than observed here.

\begin{figure*}
     \centering
     \begin{subfigure}[h]{0.49\linewidth}
         \centering
         \includegraphics[width=\textwidth]{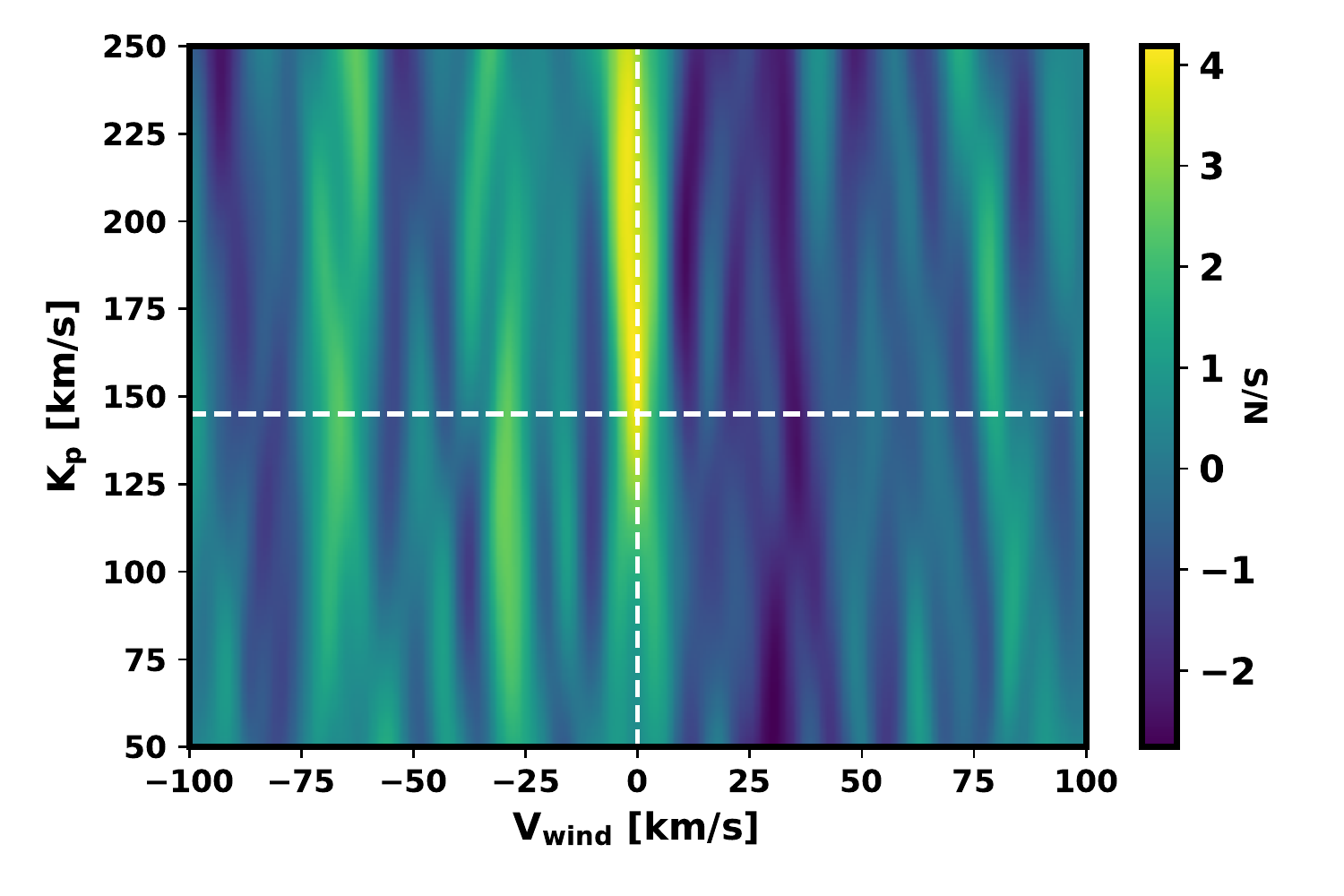}
         \caption{}
        \label{fig:hd209458b_h2o_delta_weighting}
     \end{subfigure}
     \hfill  
     \begin{subfigure}[h]{0.49\linewidth}
         \centering
         \includegraphics[width=\textwidth]{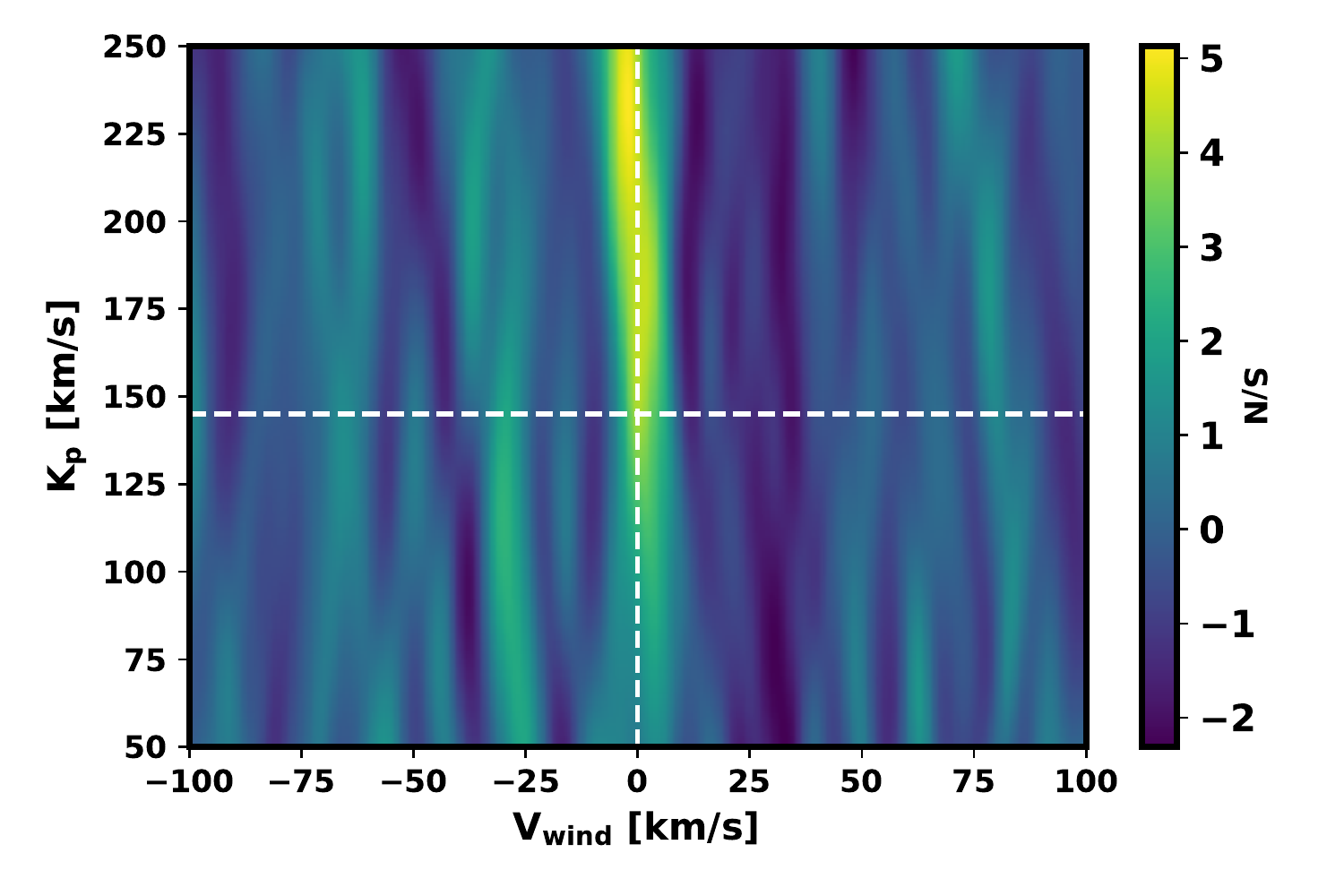}
         \caption{}
        \label{fig:hd209458b_h2o_delta_selection}
     \end{subfigure}
     \hfill    
     \begin{subfigure}[h]{0.49\linewidth}
         \centering
         \includegraphics[width=\textwidth]{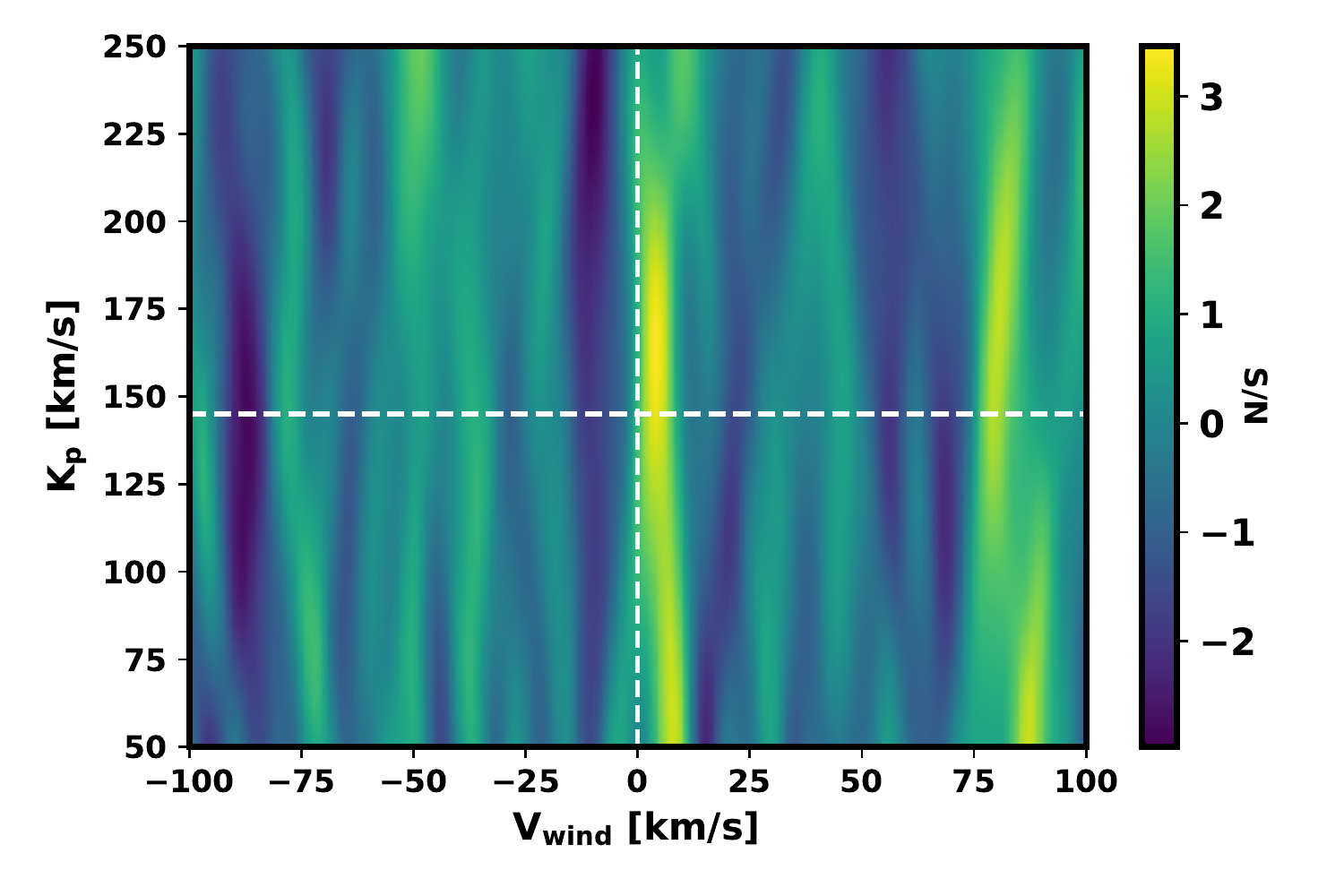}
         \caption{}
        \label{fig:hd209458b_nh3_delta_weighting}
     \end{subfigure}
     \hfill     
     \begin{subfigure}[h]{0.49\linewidth}
         \centering
         \includegraphics[width=\textwidth]{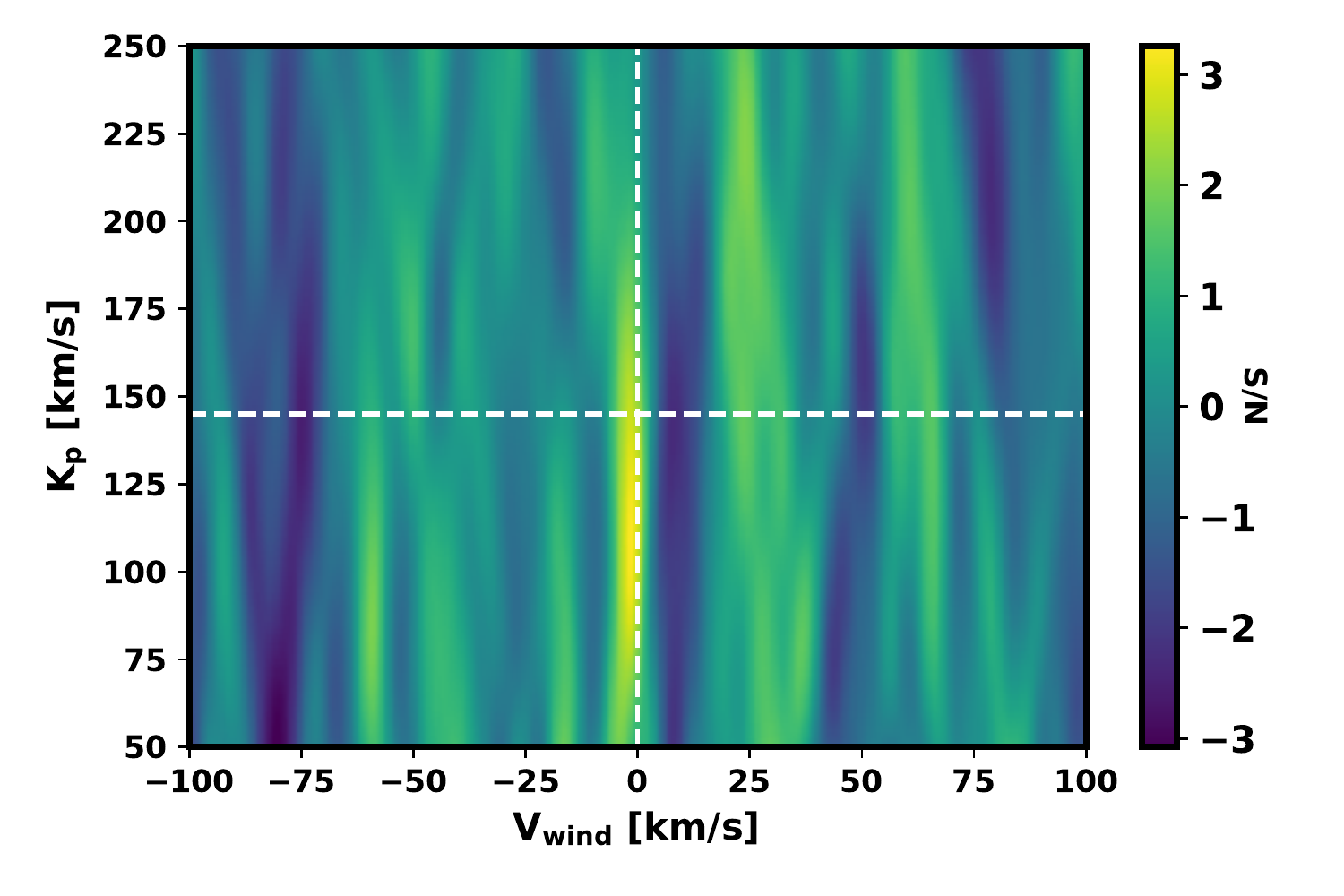}
         \caption{}
        \label{fig:hd209458b_hcn_delta_weighting}
     \end{subfigure}
 \caption{S/N maps showing non-robust signals retrieved in the atmosphere of HD 209458 b when the orders are weighted or selected according to the S/N from \ccfobsa. In each case, the number of PCA iterations applied is found by globally optimising the S/N from \deltaccf for that species. There are initially 22 orders remaining after the a priori masking. Panel (a): \ce{H2O} signal with a S/N of 4.2 obtained using order weighting. The contribution of each order to the final CCF is weighted according to its S/N from \ccfobsa, as described in Section \ref{weighting}. Panel (b): \ce{H2O} signal with a S/N of 5.1 obtained using order selection. Only the 13 best orders according to the S/N from \ccfobs are included in the final CCF. Panel (c): \ce{NH3} signal with a S/N of 3.5 obtained using order selection. Only the 14 best orders according to the S/N from \ccfobs are included in the final CCF. Panel (d): \ce{HCN} signal with a S/N of 3.3 obtained using order selection. Only the 17 best orders according to the S/N from \ccfobs are included in the final CCF.}
 \label{fig:209_order_weightings}
\end{figure*}

We now use this dataset to further demonstrate the effect of non-robustly optimising the order selection and weighting. We first detrend the spectra globally by optimising the S/N from \deltaccf for each chemical species considered in this work. We then conduct order selection and weighting, as described in Section \ref{weighting}, using the S/N from each of \deltaccf and \ccfobsa, and assess its effect on the detection significance for that species. When the orders are weighted or selected according to the S/N from \deltaccfa, the \ce{H2O} detection in Figure \ref{fig:hd209458b_h2o_delta} is preserved but we continue to retrieve no significant and robust signals for \ce{NH3} or \ce{HCN}. However, when orders are selected according to the S/N from \ccfobsa, the \ce{H2O} detection S/N is enhanced from 3.2 to 5.1, and \ce{NH3} and \ce{HCN} signals are found at S/N of 3.5 and 3.3, respectively, as shown in Figure \ref{fig:209_order_weightings}. Similar effects are seen when orders are optimally weighted rather than selected. This may demonstrate the lack of robustness in selecting or weighting orders according to the S/N from \ccfobsa, as discussed in Section \ref{weighting}.

\begin{figure}
    \centering
    \includegraphics[width=0.5\textwidth]{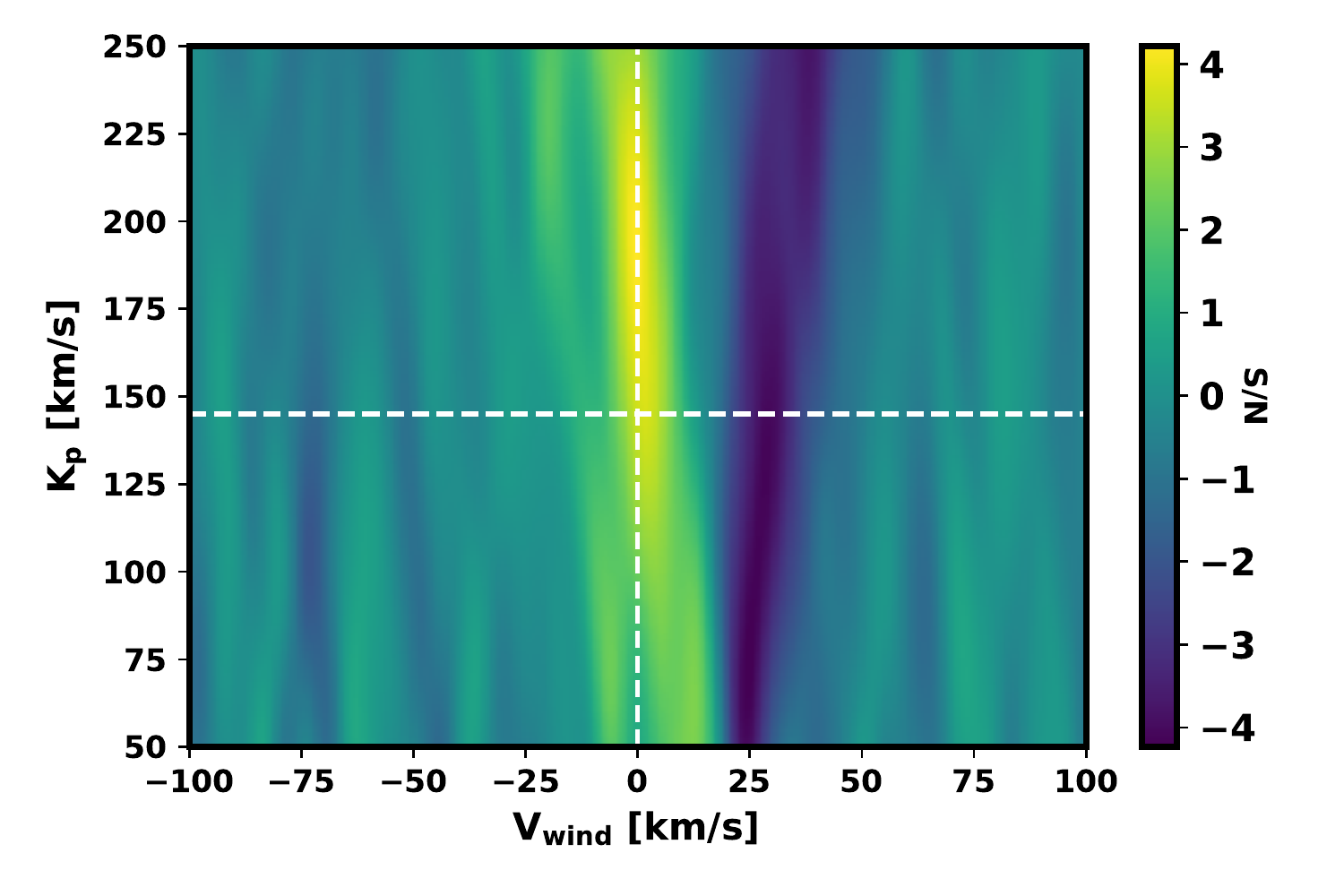}
    \caption{S/N map showing a non-robust \ce{CH4} signal (S/N = 4.2) recovered in the atmosphere of HD 209458 b when the S/N from \ccfobs is optimised order-wise in the detrending.}
    \label{fig:hd209458b_ch4_data_orderwise}
\end{figure}

As shown in Section \ref{order-wise}, order-wise optimisation of the S/N from \ccfobs during detrending may be vulnerable to the retrieval of spurious yet significant signals, and the inflation of weak signals into much more significant ones. For HD 209458 b, we retrieve a signal for \ce{CH4} with a S/N of 4.2 (Figure \ref{fig:hd209458b_ch4_data_orderwise}) when optimising order-wise the S/N from \ccfobsa, emphasising the bias induced by this detrending optimisation. This signal is completely removed when we mask the CCF within the Earth-frame velocity interval $\pm$ 5\kmsa, suggesting that is likely the result of the optimisation of uncorrected telluric residuals. 

\subsection{WASP-76 b}

We additionally analyse CARMENES archival data for a transit of the ultra-hot Jupiter WASP-76 b \citep{west2016} on the night of 4th October 2018. The data consists of 44 observations (spanning planetary orbital phases -0.0709 $< \phi <$ 0.0815), of which 25 observations are in transit. Exposure times of 498 s were used throughout, with a median observation S/N of 59. Parameters for this planet can be seen in Table \ref{wasp76b_parameters}. We use $196.5\pm0.9$ \kms \citep{ehrenreich_nightside_2020} as the expected planetary velocity. This same CARMENES dataset has been examined by a number of previous works. \cite{sanchez-lopez_searching_2022} found detections for \ce{H2O} (S/N = 5.5) and \ce{HCN} (S/N = 5.2), as well as finding an inconclusive detection of \ce{NH3} (S/N = 4.2). \cite{landman_detection_2021} meanwhile detected \ce{OH} with a S/N of 6.1.

\begin{figure}
     \centering
     \begin{subfigure}[h]{0.5\textwidth}
         \centering
         \includegraphics[width=\textwidth]{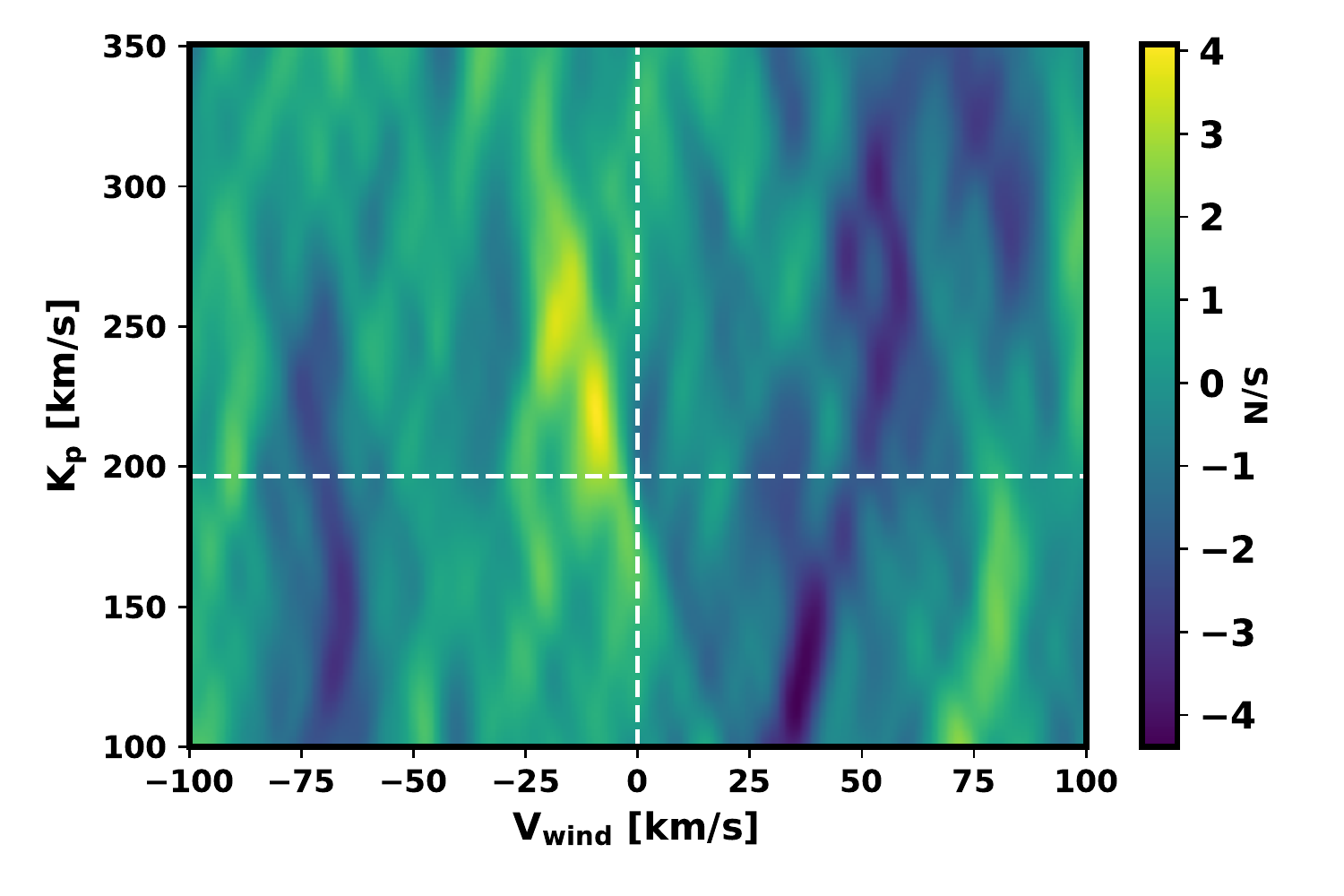}
         \caption{}
        \label{fig:wasp76b_oh_global}
     \end{subfigure}
     \hfill
     \begin{subfigure}[h]{0.5\textwidth}
         \centering
         \includegraphics[width=\textwidth]{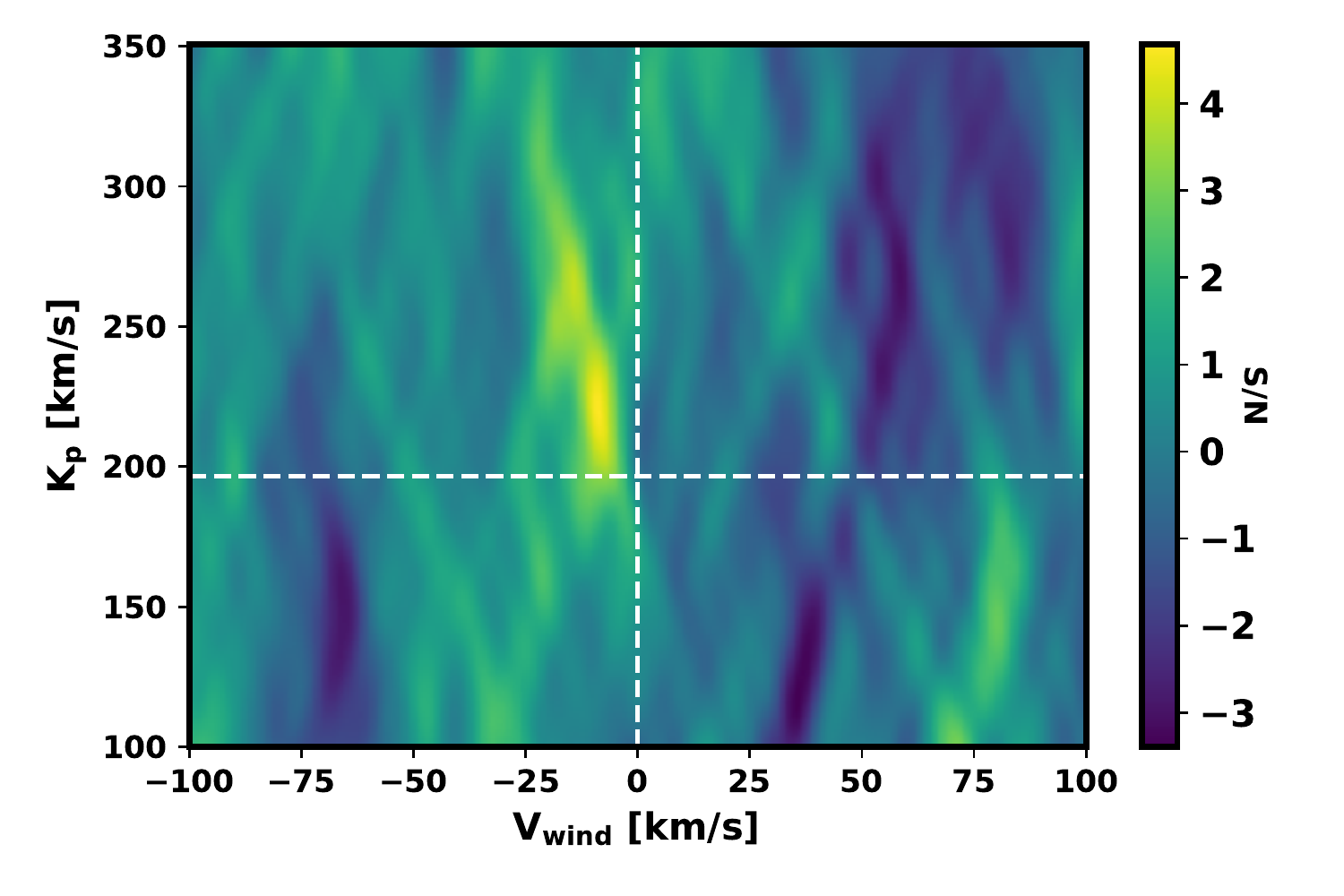}
         \caption{}
        \label{fig:wasp76_oh_orderwise}
     \end{subfigure}
 \caption{S/N maps showing \ce{OH} signals robustly found in the atmosphere of WASP-76 b. Panel (a): \ce{OH} signal with a S/N of 4.1, found by global optimisation of the S/N from \deltaccf during detrending. This corresponds to the application of 18 PCA iterations. Panel (b): \ce{OH} signal with a S/N of 4.7, found by optimising order-wise the S/N from \deltaccf during detrending.}
 \label{fig:WASP76B_OH}
\end{figure}

We remove orders 45-41 and 55-53 due to their low S/N. After observing the erosion of planetary signals synthetically injected into these spectra, as in Figures \ref{fig:signal_erosion} and \ref{fig:signal_erosion2}, we consider up to 32 PCA iterations when optimising the detrending. Since \ce{OH} is the species responsible for sky emission lines in the Earth's atmosphere, like \ce{H2O} we only consider PCA iterations of 3 or more for this species to aid the sufficient removal of emission line residuals. Optimising the S/N from \deltaccfa, first globally and then order-wise, we confirm the \cite{landman_detection_2021} detection of \ce{OH} with S/N values of 4.1 and 4.7, respectively (Figure \ref{fig:WASP76B_OH}). The position and structure of the cross-correlation signals in \kp - \vsys space could perhaps be explained by atmospheric dynamics and rotation \citep{wardenier_decomposing_2021}, if not by noise. To test if the signals spuriously arise from uncorrected sky emission line residuals, we mask the final CCF within the Earth-frame velocity interval $\pm$5 \kmsa. The OH signals are maintained, indicating that the signal is indeed likely planetary.

\begin{figure}
    \centering
    \includegraphics[width=0.5\textwidth]{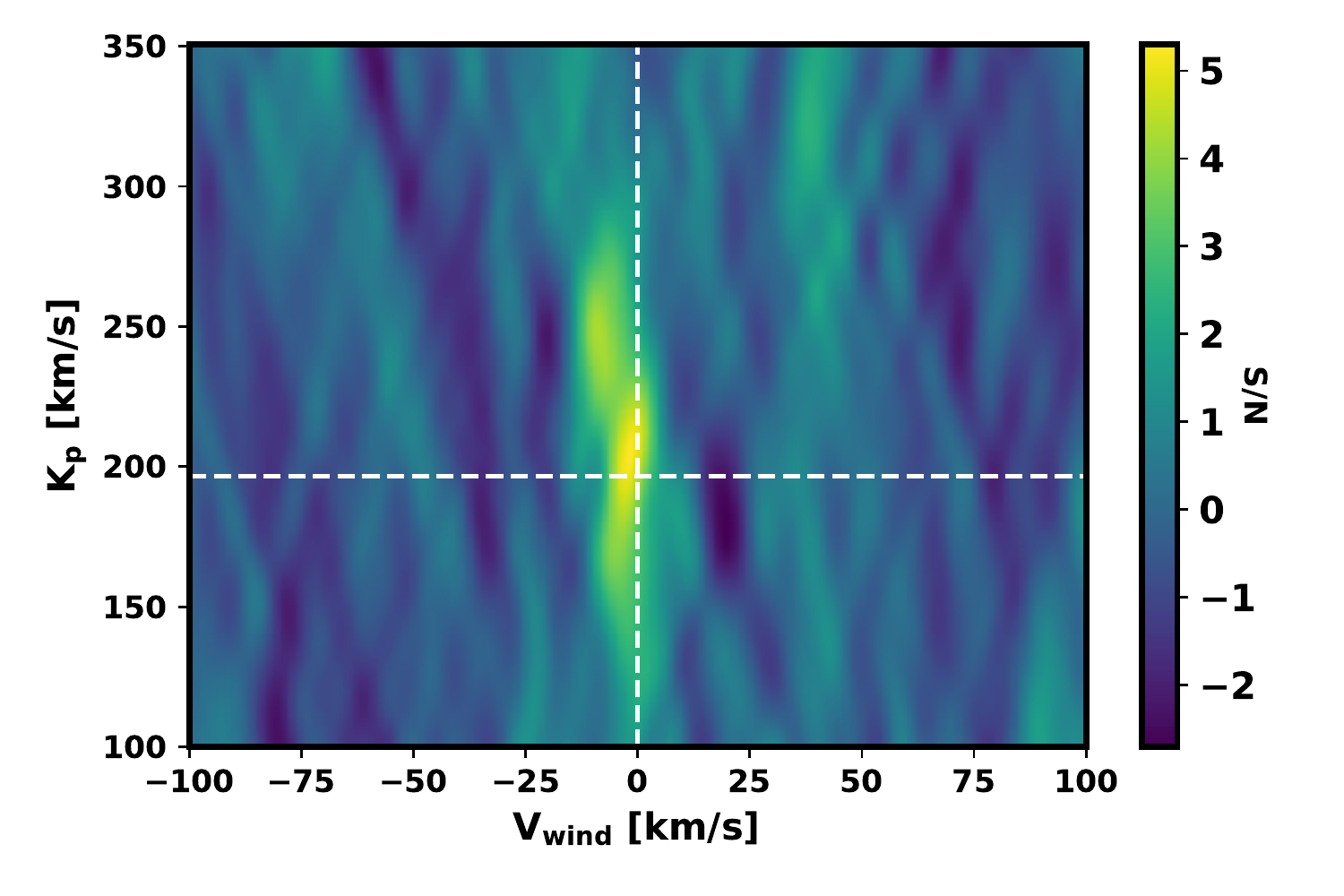}
    \caption{S/N map showing a non-robust \ce{H2O} signal recovered in the atmosphere of WASP-76 b with a S/N of 5.3. The S/N from \ccfobs is optimised order-wise in the detrending.}
    \label{fig:WASP76B_h2o}
\end{figure}

We are unable to detect any other species in this dataset using the robust methodologies established in Sections \ref{recipe} and \ref{weighting}. However, as in Section \ref{hd209458b}, we are able to obtain examples of non-robust detections for a number of species. For example, Figure \ref{fig:WASP76B_h2o} shows the recovery of an \ce{H2O} signal in the atmosphere of WASP-76 b with a S/N of 5.3, via the order-wise optimisation of the S/N from \ccfobs in the detrending. It is possible that some of the previous detections mentioned above could be reproduced robustly using a wider exploration of model space, although as discussed this is not the main goal of our work.

\section{Summary and Discussion} \label{summary}

The primary goal of this work is to investigate the robustness of methods for making chemical detections using high-resolution transmission spectra of exoplanets in the NIR. The purpose of this is twofold: to prevent false or biased detections, and to encourage consistency in such analyses across datasets. Using CARMENES observations of HD 189733 b as a case study, we examine the robustness of different PCA-based detrending optimisations, and confirm that selecting the detrending parameters to maximise the S/N of a cross-correlation signal in the presence of noise has the potential to bias detection significances at the planetary velocity of optimisation. To do this, we show that selecting detrending parameters by optimising the S/N from the direct, or `observed', CCF, \ccfobsa, can lead to detection significances which are inflated by residual noise. On the other hand, we find that optimising the S/N from the differential CCF, \deltaccfa, as defined in Section \ref{delta_method}, allows more robust detections. This appears true for both global and order-wise detrending. The robustness of optimising the S/N from the signal-injected CCF, \ccfinja, lies in between these two extremes, depending on the strength with which the model is injected into the data. As well as residual noise in \ccfinj being able to influence the selection of detrending parameters, and hence produce biased detection significances, this method is difficult to reproduce since it depends strongly on the specific models used for injection.

We also consider the robustness of weighting each spectral order's contribution to the final CCF, as is done in some previous works. Using \ccfobsa, we again demonstrate that selecting or weighting orders according to the order-wise S/N of a cross-correlation signal in the presence of noise can bias detection significances at the planetary velocity of optimisation. However, we find that such order weighting is more robust when done according to the order-wise S/N from \deltaccf at the planetary velocity of optimisation. We additionally explore how parameter choices in the analysis can influence the reported detection significance. We find that the velocity range over which we calculate the noise in the total CCF can affect the final S/N by a considerable amount (Figure \ref{fig:velocity_grid}).

We confirm a detection of \ce{H2O} in the atmosphere of HD 189733 b with a S/N of 6.1 (Figure \ref{fig:hd189733b_h2o_7}). We then conduct case studies of two further exoplanetary atmospheres, of HD 209458 b and WASP-76 b, and retrieve a signal for \ce{OH} in the atmosphere of WASP-76 b with a S/N of 4.7 (Figure \ref{fig:wasp76_oh_orderwise}). It should be reiterated that our goal is not an exhaustive exploration of model space aimed at detecting molecular species. Instead we are focused on assessing the relative robustness of molecular detections using different detrending methods for the same model template. This therefore hinders our prospects of retrieving signals for all the species that have been previously detected in our targets, and may explain why we have robustly recovered just two planetary signals across this work. It also follows that detections achieved here only via non-robust optimisations are not necessarily spurious.

Considerations in this work can be carried forward into future high-resolution spectroscopic surveys of exoplanetary atmospheres. Firstly, we have shown relative consistency in the erosion by PCA of planetary signals of different models (Figures \ref{fig:signal_erosion} and \ref{fig:signal_erosion2}), and in the detrending parameters found by optimising the S/N from \deltaccf for different models and injection velocities (Figure \ref{fig:detrending_histograms}). Therefore, it is unlikely that significant bias will be introduced due to the specific choice of injection velocity or atmospheric model, with similar detrending parameters to optimise the S/N from \deltaccf likely derived independent of such choices. Secondly, as we aim to characterise the atmospheres of smaller planets, more transits of a single target will need to be observed. The demonstrated robustness of optimising order-wise the S/N from \deltaccf could become very useful when considering observations from multiple nights; it appears robust to globally optimise the detrending of spectra from each transit individually, before combining the resultant CCF matrices. This could be important given the considerable differences in observing conditions across multiple nights, producing variable levels of telluric contamination and potentially very different optimum detrending parameters.

Other than maximising the S/N from \deltaccfa, there could be alternative methods by which to robustly optimise the number of PCA iterations during detrending. For example, one could determine the number of PCA iterations after which the residual correlated noise in \ccfobsa, e.g. from telluric, stellar and instrumental effects, has been sufficiently removed. This could be beneficial in cases where the number of PCA iterations which optimises the S/N from \deltaccf is small and coincides with there being remaining correlated noise in \ccfobsa; it is possible that this is the case in Figure \ref{fig:189_h2o_delta_orderwise}, where further detrending could potentially remove the spurious peak. Such an approach would likely require the introduction of a metric to measure the correlated noise. Alternatively, whilst this work discusses the robustness of different detrending optimisations, it would also be worthwhile to consider the efficiency of the signal extraction by considering the fraction of the planetary signal retrieved. The S/N from \deltaccf stabilises around the optimum number of PCA iterations (Figure \ref{fig:hd189733b_sn_pca}). Applying as few PCA iterations as possible to reach this stabilised S/N peak, rather than always selecting the maximum as we do here, could retain significantly more planetary information at the expense of just a small decrease in S/N. For example, whilst the S/N from \deltaccf does not change much between 3 and 13 iterations in Figure \ref{fig:hd189733b_sn_pca}, the planetary signal drops from around 75$\%$ of its initial value after 3 iterations to around 30$\%$ after 13 iterations (Figure \ref{fig:signal_erosion}). Adjusting for this could be key if we aim use the recovered planetary signal to infer more about the exoplanetary atmosphere's properties e.g. in retrievals \citep{brogi_retrieving_2019}. For robustness, a generalised method to take this into account would need to be formalised and employed homogeneously. Equally, an alternative detrending method to PCA which does not remove the planetary signal could allow for the retention of more planetary information.

Our findings motivate robust approaches for atmospheric characterisation of exoplanets using high-resolution transmission spectroscopy in the infrared. Robust and consistent methodologies would be beneficial to undertake homogeneous surveys of exoplanetary atmospheres at high spectral resolution. Inconsistencies in approaches used across different works can make it difficult to compare and contrast findings for different planets. Homogeneous surveys will allow us to place important constraints on the compositional diversity of exoplanetary atmospheres. 

\section*{Acknowledgements}
We thank the anonymous referee for the very helpful comments. This work is supported by research grants to N.M. from the MERAC Foundation, Switzerland, and the UK Science and Technology Facilities Council (STFC) Center for Doctoral Training (CDT) in Data Intensive Science at the University of Cambridge. N.M., C.C. and M.H. acknowledge support from these sources towards the doctoral studies of C.C. and M.H. We thank the CAHA Archive for making the data available. This research has made use of NASA’s Astrophysics Data System Service.

\section*{Data Availability}
This work is based on publicly available CARMENES\footnote{http://caha.sdc.cab.inta-csic.es/calto/jsp/searchform.jsp} observations.

\bibliography{Papers.bib}

\appendix
\section{Parameters}

The system parameters for HD~209458~b and WASP-76~b are shown in Tables~\ref{hd209458b_parameters} and \ref{wasp76b_parameters}, respectively.

\begin{table}
\begin{tabular}{ |p{1.0cm}||p{3.8cm}|p{2.5cm}|  }
 \hline
 Parameter & Value & Reference\\
 \hline
 P & $3.52474859\pm0.00000038$ d & \cite{knutson_using_2007}\\
 $\mathrm{T_{0}}$ & $2452826.629283\pm0.000087$ BJD & \cite{knutson_using_2007}\\
 $\mathrm{R_{star}}$ & $1.155^{+0.014}_{-0.016}$ $\mathrm{R_{\odot}}$ & \cite{torres_improved_2008}\\
 $\mathrm{R_{p}}$ & $1.359^{+0.016}_{-0.019}$ $\mathrm{R_{Jup}}$ & \cite{torres_improved_2008}\\
 \vsys & $-14.7652\pm0.0016$ \kms & \cite{mazeh_spectroscopic_2000}\\
 a & $0.04707^{+0.00046}_{-0.00047}$ au & \cite{torres_improved_2008}\\
 i & $86.71\pm0.05$ $^{\circ}$ & \cite{torres_improved_2008}\\
 $\mathrm{T_{14}}$ & $3.072\pm0.003$ hr & \cite{albrecht_obliquities_2012}\\
 \hline
\end{tabular} 
\caption{System properties of HD 209458 b, motivated by values used in \citet{sanchez-lopez_water_2019}.}
\label{hd209458b_parameters}
\end{table}

\begin{table}
\begin{tabular}{ |p{1.0cm}||p{3.8cm}| }
 \hline
 Parameter & Value\\
 \hline
 P & $1.80988198^{+0.00000064}_{-0.00000056}$ d\\
 $\mathrm{T_{0}}$ & $2458080.626165^{+0.000418}_{-0.000367}$ BJD\\
 $\mathrm{R_{star}}$ & $1.756\pm0.071$ $\mathrm{R_{\odot}}$\\
 $\mathrm{R_{p}}$ & $1.854^{+0.077}_{-0.076}$ $\mathrm{R_{Jup}}$\\
 \vsys & $-1.11\pm0.50$ \kms\\
 a & $0.0330\pm0.0002$ au\\
 i & $89.623^{+0.005}_{-0.034}$ $^{\circ}$\\
 $\mathrm{T_{14}}$ & 230 min\\
 \hline
\end{tabular}
\caption{System properties of WASP-76 b. Values adopted from \citet{ehrenreich_nightside_2020}.}
\label{wasp76b_parameters}
\end{table}

\bsp
\label{lastpage}
\end{document}